\documentclass[pdflatex,sn-nature]{sn-jnl}

\usepackage{graphicx}%
\usepackage{amsmath}
\usepackage{booktabs}

\usepackage{multirow}%
\usepackage{amsmath,amssymb,amsfonts}%
\usepackage{amsthm}%
\usepackage{mathrsfs}%
\usepackage[title]{appendix}%
\usepackage{xcolor}%
\usepackage{textcomp}%
\usepackage{manyfoot}%
\usepackage{booktabs}%
\usepackage{algorithm}%
\usepackage{algorithmicx}%
\usepackage{algpseudocode}%
\usepackage{listings}%
\usepackage{tikz}
\usepackage{tikz-cd}
\usepackage{cancel}
\usepackage{outlines}
\usepackage[version=4]{mhchem}
\usepackage{caption}

\begin{document}

\title[Article Title]{Accurate Prediction of Tensorial Spectra Using Equivariant Graph Neural Network}


\author*[1,2]{\fnm{Ting-Wei} \sur{Hsu}}\email{hsu.ting@northeastern.edu}

\author[1,2]{\fnm{Zhenyao} \sur{Fang}}\email{z.fang@northeastern.edu}

\author[1,2]{\fnm{Arun} \sur{Bansil}}\email{ar.bansil@northeastern.edu}
\author*[1,2]{and \fnm{Qimin} \sur{Yan}}\email{q.yan@northeastern.edu}


\affil[1]{\orgdiv{Department of Physics}, \orgname{Northeastern University}, \city{Boston}, \postcode{02155}, \state{Massachusetts}, \country{USA}}

\affil[2]{\orgdiv{Quantum Materials and Sensing Institute}, \orgname{Northeastern University}, \city{Burlington}, \postcode{01803}, \state{Massachusetts}, \country{USA}}


\abstract{
Optical spectroscopies provide a powerful tool for harnessing light-matter interactions for unraveling complex electronic features such as the flat bands and nontrivial topologies of materials. These insights are crucial for the development and optimization of optoelectronic devices, including solar cells, light-emitting diodes, and photodetectors, where device performance is closely connected with the nature of the underlying electronic spectrum. Realistic modeling of tensor optical responses in materials, which are computationally quite demanding, however, remains challenging. Here we introduce the Tensorial Spectra Equivariant Neural Network (TSENN), which is an equivariant graph neural network architecture that maps crystal structures directly to their full photon-frequency-dependent optical tensors. By encoding the isotropic sequential scalar components along with the anisotropic sequential tensor components into \(\ell=0\) and \(\ell=2\) spherical tensor components, TSENN ensures symmetry-aware predictions that are consistent with the constraints of crystalline symmetries of materials. Trained on a dataset of frequency-dependent dielectric tensors of 1,432 bulk semiconductors computed using first-principles methods, our model achieves a mean absolute error (MAE) of 0.127, demonstrating its potential for efficient modeling of optical properties more generally. Our framework opens new avenues for rational data--driven design of anisotropic optical responses for accelerating materials discovery for advancing optoelectronic applications.
}


\maketitle

\newpage

\section{Introduction}\label{sec1}
Optical spectroscopies, which involve excitation of a material with external electromagnetic fields, provide a powerful tool for probing properties of materials. The two key frequency-dependent response functions involved here are the dielectric function and optical conductivity, which together form the foundation for interpreting a wide range of optical phenomena in solid-state systems~\cite{fox2010optical, tanner2019optical, Lucarini2005}. Dielectric function captures the energy storage and dissipation mechanisms in the material, while optical conductivity characterizes the current response to varying external fields. Refractive index, extinction coefficient, and absorption coefficient also play important roles in specific applications such as the LEDs, integrated circuits, and solar cells~\cite{slavichExploringVanWaals2024, RUFANGURA201643, bogaertsProgrammablePhotonicCircuits2020a, hannappelIntegrationMultijunctionAbsorbers2024, marpaungIntegratedMicrowavePhotonics2019}. Optical responses are often divided into various frequency regions, where the spectra carry fingerprints of distinct types of quasiparticles in materials \cite{ZF_optical_signatures, Nagaosa:2020wc, Wang25Chiral}.

Materials with low-symmetry crystal structures exhibit pronounced anisotropies which require tensors with non-zero off-diagonal components and a basis for new strategies for controlling their optical properties. For example, direction-dependent optical responses of van der Waals heterostructures can be tailored for polarization-sensitive photodetectors and developing thermal management schemes, among other applications \cite{Liu:2016wn, Li:2023tg}. These heterostructures are also promising materials platforms for developing ultrafast lasers and artificial synaptic devices, and other transformational emerging optoelectronic technologies \cite{D0MH00340A, vdw_applications, C9NA00623K}.

Graph neural networks (GNNs) have recently shown strong capabilities in capturing structure--property relationships and efficiently predicting complex functional properties of materials~\cite{yan2024spacegroupsymmetryinformed, Kong:2022tz, Fang24GNN, Fang25GNN}. In particular, Euclidean neural networks (E(3)NN)~\cite{geiger2022e3nn, thomas2018tensorfieldnetworksrotation, cohenGroupEquivariantConvolutional2016}, which explicitly incorporate spatial symmetry constraints into their architecture can, in principle, capture the full symmetry of a crystal. This enables high-fidelity predictions with relatively small training datasets (\(\sim 10^3\) examples). E(3)NNs have demonstrated success in a variety of applications, including the prediction of interatomic potentials~\cite{batznerE3equivariantGraphNeural2022} and the construction of machine-learning Hamiltonians~\cite{gongGeneralFrameworkE3equivariant2023a, zhongTransferableEquivariantGraph2023}. 

Equivariant GNNs have also been extended to predict sequential material properties such as phonon density of states, optical conductivities, and absorption spectra~\cite{ibrahimPredictionFrequencyDependentOptical2024, Fung:2022vc, Kong:2022tz, MingdaPhonon}. For instance, \textit{GNNOpt} employs ensemble embeddings of aggregated node features to predict the trace of dielectric functions and derived properties, enabling the discovery of promising solar--cell materials and the estimation of quantum weights in materials \cite{https://doi.org/10.1002/adma.202409175}. Similarly, \textit{Optimate} and its extension were initially trained on 9,915 semiconductors and insulators using dielectric functions within the independent-particle approximation (IPA), and subsequently fine-tuned on random-phase approximation (RPA) spectra to enhance their correspondence with experimental observations~\cite{grunertDeepLearningSpectra2024, grunertMachineLearningClimbs2025}. In parallel, several works have proposed equivariant architectures for predicting tensor quantities such as dielectric, piezoelectric, and elastic tensors~\cite{heilmanEquivariantGraphNeural2024, dongAccuratePiezoelectricTensor2025}. For example, \textit{AnisoNet} leverages spherical harmonic decomposition in the \(\ell=0\) and \(\ell=2\) channels to directly predict the dielectric tensor, facilitating high-anisotropy materials discovery through database screening~\cite{lou_discovery_2025}, while \textit{DTNet} adopts a transfer-learning approach that incorporates interatomic potentials as optimized latent embeddings to construct the \(3\times3\) static dielectric constant using node features and atomic positions~\cite{maoDielectricTensorPrediction2024}. Notably, most existing studies focus exclusively on the dielectric constant. In contrast, the joint prediction of tensor structure and spectral behavior--specifically, the full frequency-dependent dielectric tensor--remains largely unexplored. Prior work has investigated tensorial spectra indirectly through equivariant machine--learning Hamiltonians, but the Brillouin-zone integrations involved continue to pose a computational bottleneck. Our approach addresses this gap by leveraging the expressive power of higher-order tensor channels within the equivariant GNNs to directly predict the full, frequency-dependent dielectric tensor from crystal structure, thereby providing a comprehensive framework for modeling anisotropic optical responses in crystalline materials.

Here we propose the Tensorial Spectra Equivariant Neural Network (TSENN) whose architecture is based on E(3)NN, which is equivariant under translation, rotation and inversion symmetries. These symmetry-aware capabilities of E(3)NN allow TSENN to model complex directional dependencies of functional properties of materials accurately. Our method utilizes sequential $\ell = 0$ components to capture isotropic parts that are associated with quantities such as the total permittivity. The sequential $\ell = 2$ components are employed to capture anisotropic parts associated with quantities such as birefringence and direction-dependent refractive index. We demonstrate the capabilities of TSENN by considering the frequency-dependent dielectric tensors of a dataset of 1,432 nonmagnetic semiconductors calculated via first-principles methods using OpenMX \cite{openmxfirst,openmxsecond, openmxthird, openmxfourth}. By training TSENN on the aforementioned dataset, we achieve a MAE of 0.127 in the predicted imaginary part of the dielectric tensor. The high prediction accuracy enables reconstruction of the real part of the dielectric tensor via the Kramers-Kronig (K-K) relations, allowing us to compute additional optical properties derived from the spectra. Accurately predicting the full tensor while preserving its symmetry opens new opportunities for designing materials with tailored anisotropic responses, with applications in photovoltaics, optical modulators, and sensing technologies.

\section{Results}\label{sec2}
\subsection{Dataset}
With our focus on small bandgap semiconductors in mind, we selected materials from the Materials project~\cite{jainCommentaryMaterialsProject2013} with band gaps ranging from 0.3$-$3 eV (photon wavelengths of 4,100~nm (infrared)  to 413~nm (violet)) that are appropriate for optoelectronic applications \cite{kangComputationalScreeningIndirectGap2019, yang2022highthroughputopticalabsorptionspectra}. To maintain structural stability, we imposed an energy-above-hull threshold of 0.02 eV/atom. For computational efficiency, we limited our selection to materials with fewer than three elements and fewer than ten atoms per unit cell. Additionally, we excluded magnetic materials as well as materials containing elements with \(f\) electrons due to the lack of availability of accurate predefined pseudo-potential sets. These constraints yielded a dataset of 1,432 materials. Importantly, we did not impose symmetry constraints, such as filtering for specific space groups. As a result, our dataset includes anisotropic materials, whose tensor properties are not intrinsically diagonal. For each entry, we first symmetrized the structure by following the procedure of Ref. \cite{SETYAWAN2010299} and performed first-principles calculations using OpenMX. The dielectric tensors were computed using the Kubo formula for photon energies ranging from 0$-$30~eV, with an increment of 0.01~eV, resulting in 1,432 frequency-dependent dielectric tensors satisfying the symmetry constraints of the crystal systems considered~\cite{bir1974chapter5,boyd2008susceptibility, powell2010chapter3}. 
In this work, the complex dielectric tensor refers to the frequency-dependent relative dielectric tensor, expressed as
$\varepsilon^{\alpha\beta}(\omega) = \varepsilon^{\alpha\beta}_1(\omega) + i\varepsilon^{\alpha\beta}_2(\omega)$,
where $\varepsilon^{\alpha\beta}_1$ and $\varepsilon^{\alpha\beta}_2$ denote the real and imaginary parts, respectively. The superscripts $\alpha\beta$ index the directional dependence of the spectra. For simplicity, we omit the explicit frequency dependence $(\omega)$ hereafter.

\begin{figure}[H]
    \centering
     \includegraphics[width=0.8\linewidth]{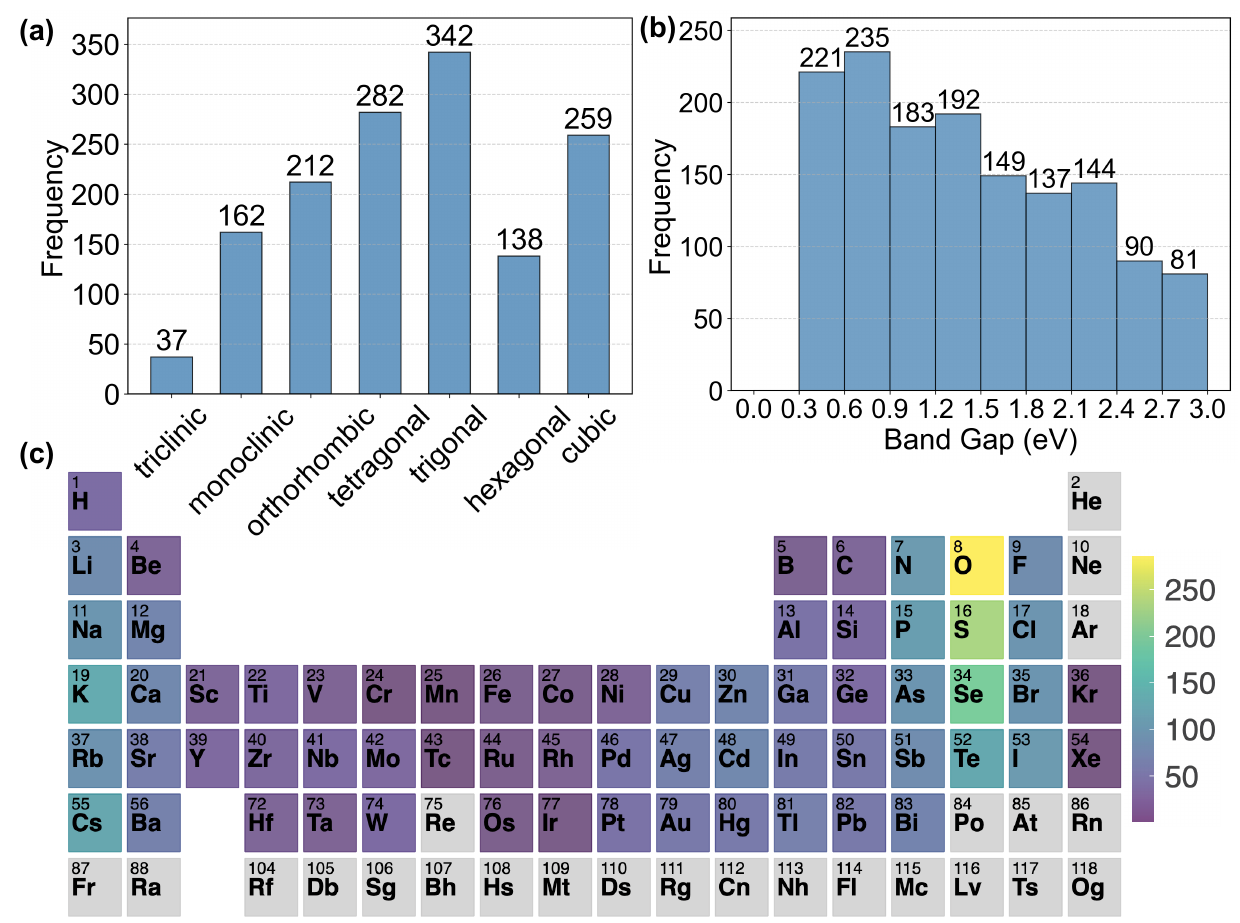}
    \caption{\textbf{Overview of the diversity of materials in our dataset.} (a) Histogram showing the frequency of materials across different crystal symmetries (b) Distribution of band gaps (eV). (c) Periodic table highlighting the frequency of elements in the dataset, with the color gradient ranging from blue (low frequency) to yellow (high frequency). Lanthanide and Actinide series are filtered out and not displayed.}
    \label{fig:1}
\end{figure}

\begin{figure}[H]
    \centering
    \includegraphics[width=0.8\linewidth]{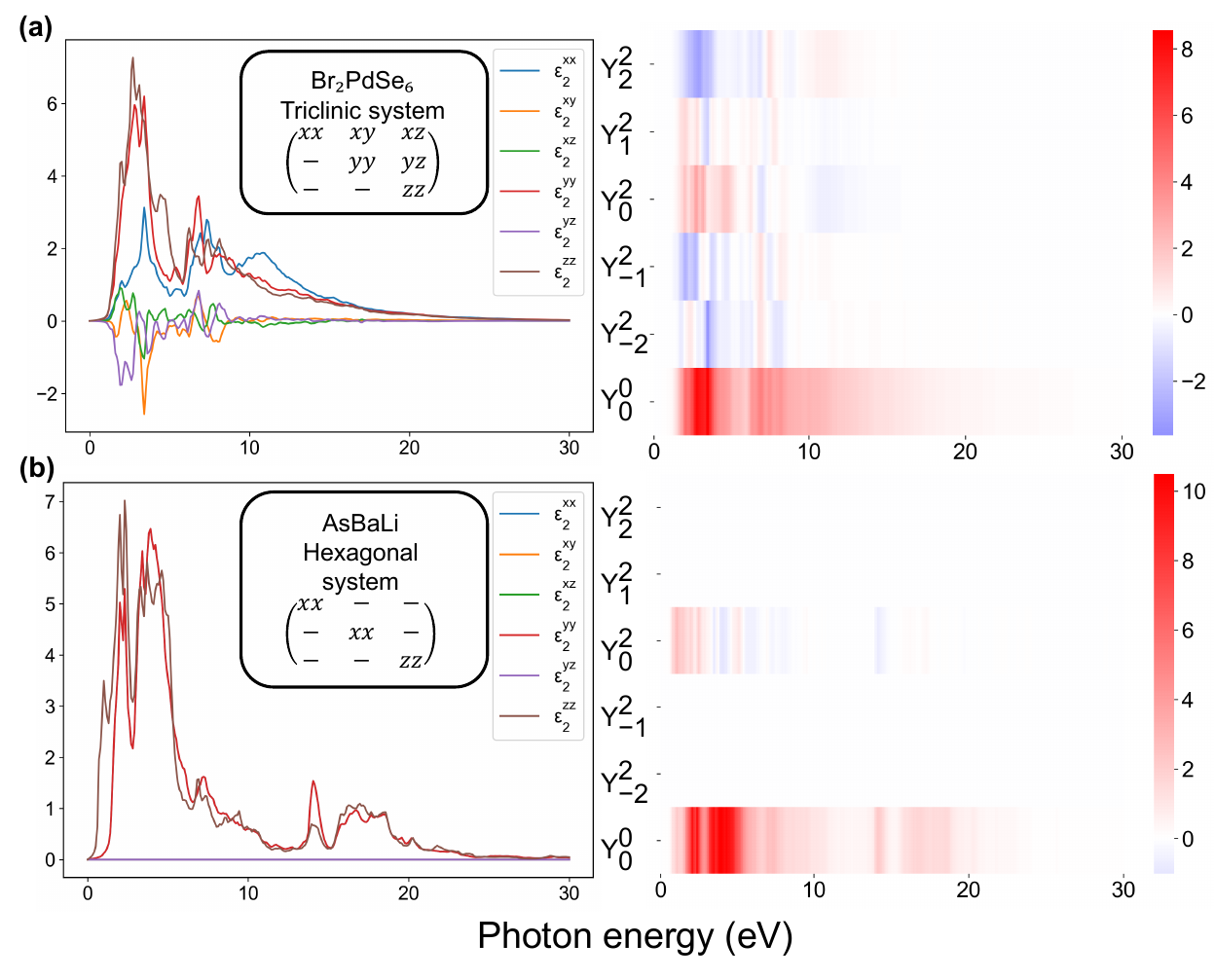}
    \caption{\textbf{Dataset visualization.} Left panels illustrate the frequency-dependent dielectric tensors in Cartesian coordinates, with insets detailing the corresponding crystal systems, tensor formulas, and the non-vanishing components after Ref.~\cite{powell2010chapter3}. Right panels depict the transformed spectra into the spherical harmonic basis, serving as the ground truth for model training. (\textbf{a}) Triclinic system with all tensor components present. (\textbf{b}) Hexagonal system with non-vanishing components satisfying \(xx = yy \neq zz\), where the \(Y^0_0\) component arises from the \(xx + yy + zz\) component and the \(Y^2_0\) arises from the \(-\frac{1}{2}(xx + yy) + zz\) component.}
    \label{fig:spherical_harmonics}
\end{figure}

To predict the dielectric tensor using an equivariant neural network, we express it in the spherical--harmonics basis, which naturally encodes rotational symmetry. The dielectric tensor is symmetric under the permutation of Cartesian indices, $\varepsilon^{ij} = \varepsilon^{ji}$, and therefore admits the decomposition
\begin{equation}\label{eq:change_of_basis}
\varepsilon^{ij} = T^{\alpha\beta} \, Y_1^\alpha Y_1^\beta \;=\; \varepsilon^{(0)} \oplus \varepsilon^{(2)} \;=\; \sum_{\ell \in \{0,2\}} \sum_{m=-\ell}^{\ell} \varepsilon_{(\ell)}^{m} \, Y_{\ell}^{m},
\end{equation}
which separates the scalar trace component ($\ell = 0$) from the traceless symmetric tensor component ($\ell = 2$). Details of the decomposition procedure are provided in SI Sec.~6. This approach transforms the tensors from the Cartesian basis into the spherical--harmonics basis, making it easier to work with tensor properties in systems with rotational and inversion symmetries. In Eq.~\eqref{eq:change_of_basis}, note that a rank-2 Cartesian tensor can be decomposed into \(\ell=0\) and \(\ell=2\) channels, which involve \(2\ell + 1\) different magnetic quantum numbers \(m \in \{-\ell, \ldots, \ell\}\) that form a spherical tensor of order \(\ell\). Since we are dealing with frequency-dependent tensors, we take the Cartesian tensor at each photon energy, perform this change-of-basis operation, and obtain the frequency-dependent spherical harmonic coefficients as our target as shown in Fig.~\ref{fig:spherical_harmonics}. The change of basis and the evaluation of the Wigner \(3j\) coefficients were carried out using the \texttt{e3nn} package~\cite{geiger2022e3nn}.

Fig.~\ref{fig:spherical_harmonics} illustrates the spherical--harmonics decomposition of the dielectric tensor for two representative materials from our dataset, which are shown both in the Cartesian and spherical--harmonics basis. It can be shown through a crystallographic analysis that there are five distinct cases of non-vanishing dielectric tensor components across the seven crystal systems. We highlight two examples. (i) Triclinic compound \ce{Br2PdSe6}, which lacks any symmetry operations, so that all components of the dielectric tensor are non-vanishing. Consequently, the decomposition of its Cartesian dielectric tensor yields non-zero contributions across all irreducible spherical components, including the scalar channel \( Y_0^0 \) and the five rank-2 spherical harmonic components \( Y_m^2 \) for \( m \in \{-2, -1, 0, 1, 2\} \). And (ii) the hexagonal compound \ce{AsBaLi} in which the higher crystal symmetry imposes constraints on the tensor. Specifically, the sixfold rotational symmetry enforces the presence of diagonal components only with the relationships \( \varepsilon^{xx} = \varepsilon^{yy} \neq \varepsilon^{zz} \). The \( xx \) and \( yy \) components (blue and red curves in Fig.~\ref{fig:spherical_harmonics}) are visually indistinguishable within numerical accuracy. All off-diagonal components vanish, consistent with the symmetry constraints summarized in the inset of Fig.~\ref{fig:spherical_harmonics}. As a result, the spherical--harmonics decomposition becomes  sparse, with only the \( Y_0^0 \) and \( Y_0^2 \) components remaining nonzero. Origin of these spherical channels can be traced back to the specific linear combinations of Cartesian components: the scalar component \( Y_0^0 \) arises from the trace \( {xx} + {yy} + {zz} \), while the rank-2 spherical harmonic components correspond to the combination \( -\frac{1}{2} ({xx} + {yy}) + {zz} \) \cite{mochizukiSphericalHarmonicDecomposition1988a}. 

\subsection{Equivariant Graph Neural Network Architecture}
\begin{figure}[H]
    \centering
    \includegraphics[width=0.8\linewidth]{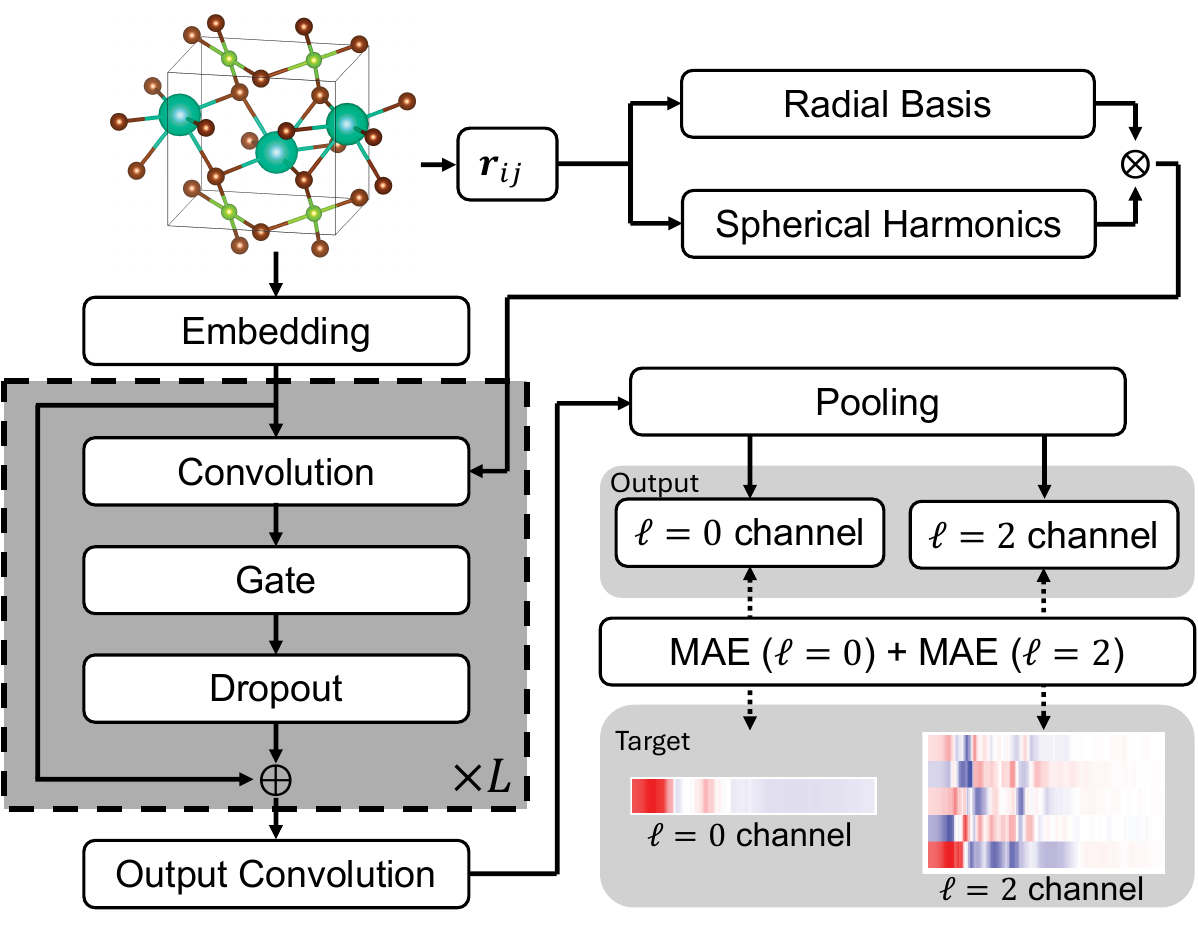}
    \caption{
    Schematic architecture of TSENN. The model takes a periodic crystal graph as input and encodes atom types via one-hot embeddings \cite{ xieCrystalGraphConvolutional2018a}. Edge vectors $\mathbf{r}_{ij}$ are expanded using radial basis functions and spherical--harmonics. Features pass through multiple gated equivariant convolution layers, followed by output convolution and mean pooling. Dropout is incorporated after each convolution block to prevent overfitting. The prediction is split into $\ell=0$ (isotropic) and $\ell=2$ (anisotropic) channels and trained using the composite loss function: $\mathrm{MAE}(\ell=0) + \mathrm{MAE}(\ell=2)$.}
    \label{fig:model}
\end{figure}

Our TSENN model extends the \texttt{e3nn} and GNNOpt frameworks, both of which have demonstrated strong performance in sequential prediction tasks such as electron and phonon densities of states, as well as the traces of optical spectra~\cite{MingdaPhonon,https://doi.org/10.1002/adma.202409175,okabe2024virtual}. While TSENN is built on the foundational principles of equivariant GNNs as implemented in \texttt{e3nn}, it incorporates key extensions that distinguish it from GNNOpt. In particular, TSENN leverages higher-order spherical harmonic components to better capture the anisotropy of tensorial properties (Figure~\ref{fig:model}). The effectiveness of the output representation, model architecture, and comparison with state-of-the-art methods is further detailed in the Supplementary Information (SI). Accordingly, we construct a GNN that represents each material as a crystal graph, where the atomic and structural information is encoded through node and edge features. Each atomic node \( i \) is initially represented by a one-hot encoding of its atomic species \(z_{\text{type}, i} = \text{onehot}(\text{atomic number}) \in \mathbb{R}^{118},\)
corresponding to the 118 known chemical elements~\cite{xieCrystalGraphConvolutional2018a}. These atomic types are embedded via a multilayer perceptron to produce initial node-level features \( f_i \). 

TSENN enforces equivariance to 3D Euclidean transformations by adopting the formalism of tensor field networks~\cite{thomas2018tensorfieldnetworksrotation}. Translational equivariance is achieved through the use of relative atomic positions in message passing, while rotational equivariance is enforced by constraining intermediate features to lie within irreducible representations of the rotation group \(\mathrm{SO}(3)\). These are indexed by rotation order \(\ell \in \{0, 1, 2, \ldots\}\) and parity \(p \in \{-1, 1\}\). Each feature tensor takes the form \(V_{cm}^{(\ell, p)}\), where \(c\) indexes feature channels and \(m \in [-\ell, \ell]\) indexes spherical harmonic components. Thus, a representation of degree \(\ell\) occupies a tensor space of dimension \(\mathbb{R}^{N_{\text{channels}} \times (2\ell + 1)}\).

These representations are combined using equivariant tensor product operations \(\otimes\) to construct the pointwise convolution layers. This operation can be applied iteratively \(L\) times, following the Clebsch--Gordan decomposition as implemented in \texttt{e3nn}:
\begin{equation}
\left(\mathbf{U}^{(\ell_1, p_1)} \otimes \mathbf{V}^{(\ell_2, p_2)}\right)^{(\ell_0, p_0)}_{c m_0} = 
\sum_{m_1 = -\ell_1}^{\ell_1} \sum_{m_2 = -\ell_2}^{\ell_2}
C^{(\ell_0, m_0)}_{(\ell_1, m_1)(\ell_2, m_2)} \,
U^{(\ell_1, p_1)}_{c m_1} \,
V^{(\ell_2, p_2)}_{c m_2},
\end{equation}
where \(\ell_0 \in \left[|\ell_1 - \ell_2|, \ell_1 + \ell_2\right]\) and \(p_0 = p_1 p_2\). For brevity, we omit the indices \(\ell\), \(p\), \(c\), and \(m\) when referring to these tensors hereafter.

After the final pointwise convolution layer, we apply mean pooling over all nodes to obtain a global representation of the material. The network output is then partitioned into two parts: a scalar component corresponding to \(\ell = 0\) and a tensor component corresponding to \(\ell = 2\). To train the model, we define a composite loss function \(\mathcal{L}\), which separately evaluates the scalar and tensor components of the predicted optical spectra. Let \(N\) denote the number of materials and \(n_\omega\) the number of frequency points. For each material entry \(i\), \(\hat{y}_{0e}^{(i)}(\omega)\) and \(\hat{y}_{2e}^{(i)}(\omega, m)\) denote the predicted \(\ell = 0\) and \(\ell = 2\) components, respectively, and \(y_{0e}^{(i)}(\omega)\) and \(y_{2e}^{(i)}(\omega, m)\) denote the corresponding ground-truth values. Here, the hat symbol denotes the predicted tensor, unless explicitly specified otherwise, distinguishing it from the ground-truth tensor obtained via the Kubo formula. Since there are two objectives ($\mathcal{L}_1 (\ell=0)$ and $\mathcal{L}_2 (\ell=2)$) with different magnitudes, we adopt the loss balancing approach for multi-objective training, which allows the network to determine the optimal weighting between objectives dynamically \cite{kendallMultiTaskLearningUsing2018}. The effectiveness of this loss balancing is discussed in SI Sec.~2.2. The total loss is defined as:
\begin{equation}
\mathcal{L}=\frac{1}{2 \sigma_1^2} \mathcal{L}_1(\ell=0)+\frac{1}{2 \sigma_2^2} \mathcal{L}_2(\ell=2)+\log \sigma_1+\log \sigma_2
\end{equation}
where \(\sigma_1\) and \(\sigma_2\) are learnable parameters that represent the relative uncertainty of each task, with component-wise  definitions:
\begin{equation}
\mathcal{L}_0 (\ell=0) = \frac{1}{N n_\omega} \sum_{i=1}^{N} \sum_{\omega } | \hat{y}_{0e}^{(i)}(\omega) - y_{0e}^{(i)}(\omega) |,
\end{equation}

\begin{equation}
\mathcal{L}_2 (\ell=2) = \frac{1}{N n_\omega M_{\ell = 2}} \sum_{i=1}^{N} \sum_{\omega} \sum_{m = -2}^{2} | \hat{y}_{2e}^{(i)}(\omega, m) - y_{2e}^{(i)}(\omega, m) |,
\end{equation}
where \(M_{\ell = 2} = 2 \cdot 2 + 1 = 5\) is the number of components for the rank-2 irreducible representation. The model used for prediction was selected based on the best performance according to the validation loss. After obtaining the predicted spherical tensor components from the best model, we transformed them back into the Cartesian tensor representation. 

To evaluate the model's performance, the most direct measure is the mean absolute error (MAE) of the dielectric tensor,
\begin{equation}
\text{MAE}=\frac{1}{6 N_\omega} \sum_\omega \sum_{\alpha \leq \beta}\left|\hat \varepsilon^{\alpha \beta}(\omega)-\varepsilon^{\alpha \beta}(\omega)\right|.
\end{equation}
This ensures that peak position, width, and sign are faithfully captured in the evaluation, without cancellation across frequencies or components. In addition, we compute the per--component errors $\text{MAE}^{\alpha\beta}$, as well as a Normalized MAE ($\text{NMAE}^{\alpha\beta}$), defined by normalizing $\text{MAE}^{\alpha\beta}$ to the characteristic magnitude of each tensor component. This scale--aware metric highlights how the prediction errors compare to the intrinsic strength of the typical dielectric response, thereby providing additional physical interpretability.

To isolate anisotropic behavior, we remove the isotropic contribution from the tensor,
\begin{equation}
\varepsilon_{\text{aniso}}(\omega)=\varepsilon(\omega)-\frac{1}{3}\operatorname{Tr}[\varepsilon(\omega)] I.
\end{equation}
This decomposition yields the component--wise anisotropic contributions $\varepsilon_{\text{aniso}}^{\alpha \beta}(\omega)$ and their overall magnitudes $\left\|\varepsilon_{\text{aniso}}(\omega)\right\|_F$. From these, we define per–component errors $\text{MAE}_{\text{aniso}}^{\alpha\beta}$, the related component--averaged value $\text{MAE}^{\text{aniso}}$, and the Frobenius--norm errors $\text{MAE}_{\text{aniso}}^{\text{norm}}$, to quantify, errors in the overall anisotropy strength. For completeness, we also report additional shape-sensitive metrics such as the first--derivative MAE of individual components $\mathrm{MAE}^{\prime, \alpha \beta}$ and the KL divergence for the diagonal components. Detailed definitions of these metrics are provided in the SI Sec.~1.

\subsection{Prediction Results}
Table~\ref{tab:dielectric_errors} summarizes error metrics for dielectric tensor predictions across chemically and structurally diverse test systems. The overall tensor MAE is 0.127 (0.097 median), establishing TSENN's robust baseline accuracy. When decomposed into isotropic and anisotropic contributions, the averaged anisotropic error $\text{MAE}_{\text{aniso}}$ is 0.096 (0.074 median), demonstrating reliable capture of directional variations. By contrast, the Frobenius–norm anisotropy error $\text{MAE}_{\text{aniso}}^{\text{norm}}$ is larger at 0.141 (0.111 median), reflecting the stricter requirement of reproducing the total anisotropy magnitude rather than individual Cartesian components. We emphasize that this norm-based measure is the most physically meaningful measure, since it is rotationally-invariant and directly corresponds to the $\ell=2$ spherical--harmonics channel. This invariance complements the model's equivariance, which guarantees the correct transformation of anisotropic components as a rank-2 tensor under rotations.

At the component level, the diagonal elements ($xx$, $yy$, $zz$) exhibit higher errors (0.237--0.259, with NMAE $\sim$3\%) compared to the off-diagonal terms ($xy$, $xz$, $yz$) (0.082--0.105, with NMAE $\sim$5--7\%). This hierarchy is physically natural: the diagonal responses are much stronger in magnitude and receive contributions from both the isotropic ($\ell=0$) and anisotropic ($\ell=2$) channels, whereas the off-diagonal terms arise purely from the $\ell=2$ sector and vanish in many high-symmetry systems. Note that we do not apply per-component normalization or weighting: equivariance permits only blockwise scalings that are constant on each SO(3) irreducible subspace. Any differential weighting inside the $\ell=2$ block
(e.g., different weights for $m=\pm 2,\pm 1,0$ or per--Cartesian rescaling) would break rotational equivariance.

Beyond absolute and relative magnitudes, values of first-derivative MAEs (0.094--0.105 for diagonal and 0.034--0.039 for off-diagonal terms) and KL divergences (0.106–0.161 for diagonal term) confirm that the model reproduces not only overall intensities but also the line shapes and peak positions. The close agreement between the mean and median values across all metrics further indicates that performance is uniform across diverse systems, without being skewed by outliers. These results demonstrate that TSENN achieves high fidelity in predicting the full dielectric tensor, capturing both the dominant diagonal responses and the more subtle anisotropic features with accuracy and physical consistency.

\begin{figure}[H]
    \centering
    \includegraphics[width=0.98\linewidth]{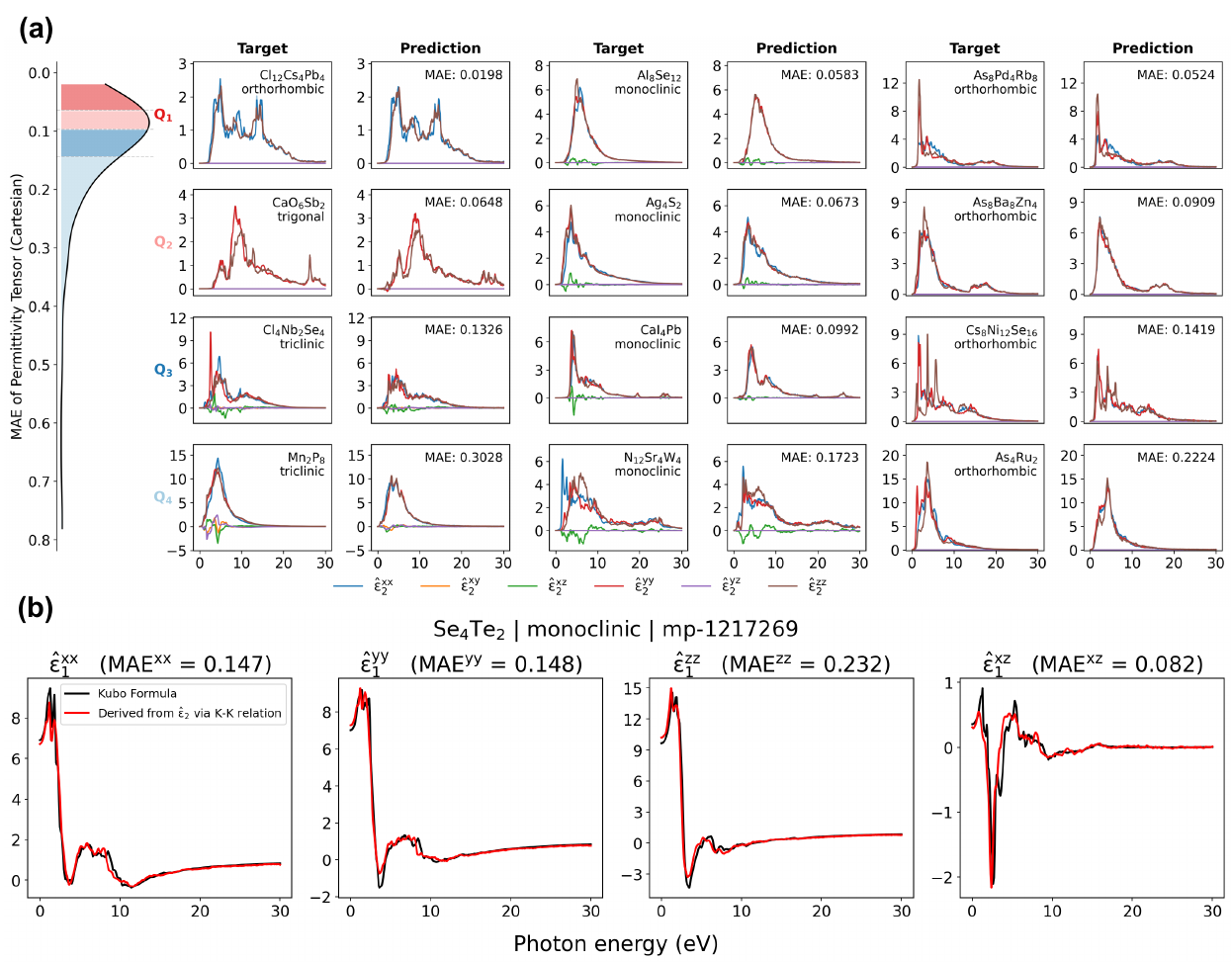}
    \caption{\textbf{(a)} Distribution of full-tensor MAE values across the test set (left) and representative spectra obtained by stratified sampling across quartiles (right). From the cumulative kernel-density-estimator (KDE) plot, we randomly selected three systems in each quartile ($Q_1$--$Q_4$), resulting in 12 representative examples that span the full error distribution. [Cubic lattices were excluded, as their isotropic spectra are already well studied.] For each sampled material, the target and predicted tensor components are shown side by side, with the corresponding MAE reported in each panel. This stratified sampling strategy ensures that both the high- and low-error cases are represented, providing a balanced and unbiased view of model's performance across crystal systems and anisotropy strengths.
    \textbf{(b)} Real parts of the dielectric tensor components for the monoclinic system \ce{Se4Te2}. 
Black curves give the \textit{ab initio} dielectric spectra based on the IPA using the Kubo formula, 
while red curves denote spectra reconstructed from the predicted imaginary part via the K-K relation. 
$\text{MAE}^{\alpha\beta}$ is reported in each panel, with values ranging from 0.082 to 0.232, demonstrating strong agreement across all non-vanishing components 
(\(\varepsilon^{xx}, \varepsilon^{yy}, \varepsilon^{zz}, \varepsilon^{xz}\)). 
Notably, both the off-diagonal element (\(\varepsilon^{xz}\)) and the anisotropic diagonal terms are faithfully recovered.
    }
    \label{fig:prediction_results}
\end{figure}

To visualize the predictions, we stratified the test set based on MAE values and selected representative materials from each quartile, excluding cubic lattices since their isotropic dielectric response is already well understood. This sampling highlights systems with nontrivial anisotropies [Fig.~\ref{fig:prediction_results}(a)].
With the accurate predicted imaginary parts, we used K-K relations to obtain the real parts of the dielectric tensors. Since K-K relations are model-independent, they provide a rigorous test of the internal consistency of the predicted spectra. In Fig.~\ref{fig:prediction_results}(b), we compare the real parts obtained via K-K relations from the predicted imaginary spectra against the reference real parts computed directly using the Kubo formula. An exemplar monoclinic system in the test set is used (\ce{Se4Te2}), where due to symmetry constraints, only four components of the dielectric tensor \(\hat{\varepsilon}^{ij}\) are non-zero: \( \hat{\varepsilon}^{xx}_2\), \( \hat{\varepsilon}^{yy}_2\), \( \hat{\varepsilon}^{zz}_2\), and \( \hat{\varepsilon}^{xz}_2\), while the off-diagonal components \( \hat{\varepsilon}^{xy}_2\) and \( \hat{\varepsilon}^{yz}_2\) vanish. As shown in the figure, the reconstructed components exhibit excellent agreement with the Kubo-derived results, demonstrating that both the off-diagonal elements (e.g., $\hat\varepsilon_1^{xz}$) and the anisotropic diagonal elements ($\hat\varepsilon_1^{xx}$, $\hat\varepsilon_1^{yy}$, $\hat\varepsilon_1^{zz}$) are faithfully reproduced. The other off-diagonal components, which are not shown, are found to be negligible in keeping with symmetry constraints. This analysis validates not only the predictive accuracy of the imaginary dielectric spectrum but also the physical plausibility and self-consistency of the full complex tensor. These results demonstrate the model's potential for downstream applications, including photonic materials discovery, symmetry-aware property screening, and the inverse design of materials with tailored optical anisotropy.

\begin{table}[ht]
\captionsetup{labelformat=empty} 
\caption{\textbf{Table 1.} Summary of full-tensor and per-component error metrics for dielectric tensor predictions (mean values with median in parentheses).}
\label{tab:dielectric_errors}
\begin{tabular*}{\textwidth}{@{\extracolsep\fill}cc|cccccc}
\toprule
\multicolumn{2}{c|}{\textbf{Full-tensor metrics}} & \multicolumn{5}{c}{\textbf{Per-component metrics }} \\
\midrule
Metric & Value & $\alpha\beta$ & $\mathrm{MAE}^{\alpha \beta}$ & $\mathrm{NMAE}^{\alpha \beta}$ (\%)  & $\mathrm{MAE}^{\prime, \alpha \beta}$ & KL divergence & $\text{MAE}_{\text{aniso}}^{\alpha\beta}$ \\
\midrule
MAE                         & 0.127 (0.097) & $xx$ & 0.251 (0.176) & 2.9 (2.8)& 0.099 (0.084) & 0.119 (0.047) & 0.092 (0.066) \\
$\text{MAE}_{\text{aniso}}$ & 0.096 (0.074) & $yy$ & 0.237 (0.177) & 2.8 (2.9)& 0.094 (0.082) & 0.106 (0.047) & 0.083 (0.064) \\
$\text{MAE}_{\text{aniso}}^{\text{norm}}$ & 0.141 (0.111) & $zz$ & 0.259 (0.206)& 3.0 (3.2) & 0.105 (0.092) & 0.161 (0.067) & 0.125 (0.099) \\
\cmidrule(lr){3-8}
 & & $xy^\dagger$ & 0.082 (0.072) & 5.4 (8.4) & 0.034 (0.032) & -- & 0.082 (0.072) \\
 & & $xz^\dagger$ & 0.091 (0.089) & 7.3 (7.4) & 0.037 (0.034) & -- & 0.091 (0.089) \\
 & & $yz^\dagger$ & 0.105 (0.098) & 7.0 (8.0) & 0.039 (0.039) & -- & 0.105 (0.098) \\
\botrule
\end{tabular*}
\begin{flushleft}
\footnotesize{$^\dagger$ Off-diagonal values are reported only for symmetry-allowed systems. 
Since the isotropic trace contributes only to diagonal elements, 
$\text{MAE}^{\alpha\beta} = \text{MAE}_{\text{aniso}}^{\alpha\beta}$ for off-diagonal terms. 
The KL divergence is not reported for off-diagonal components because these spectra are not positive definite, 
and the normalization required for KL is ill-defined.}
\end{flushleft}
\end{table}

\begin{figure}[H]
    \centering
    \includegraphics[width=0.9\linewidth]{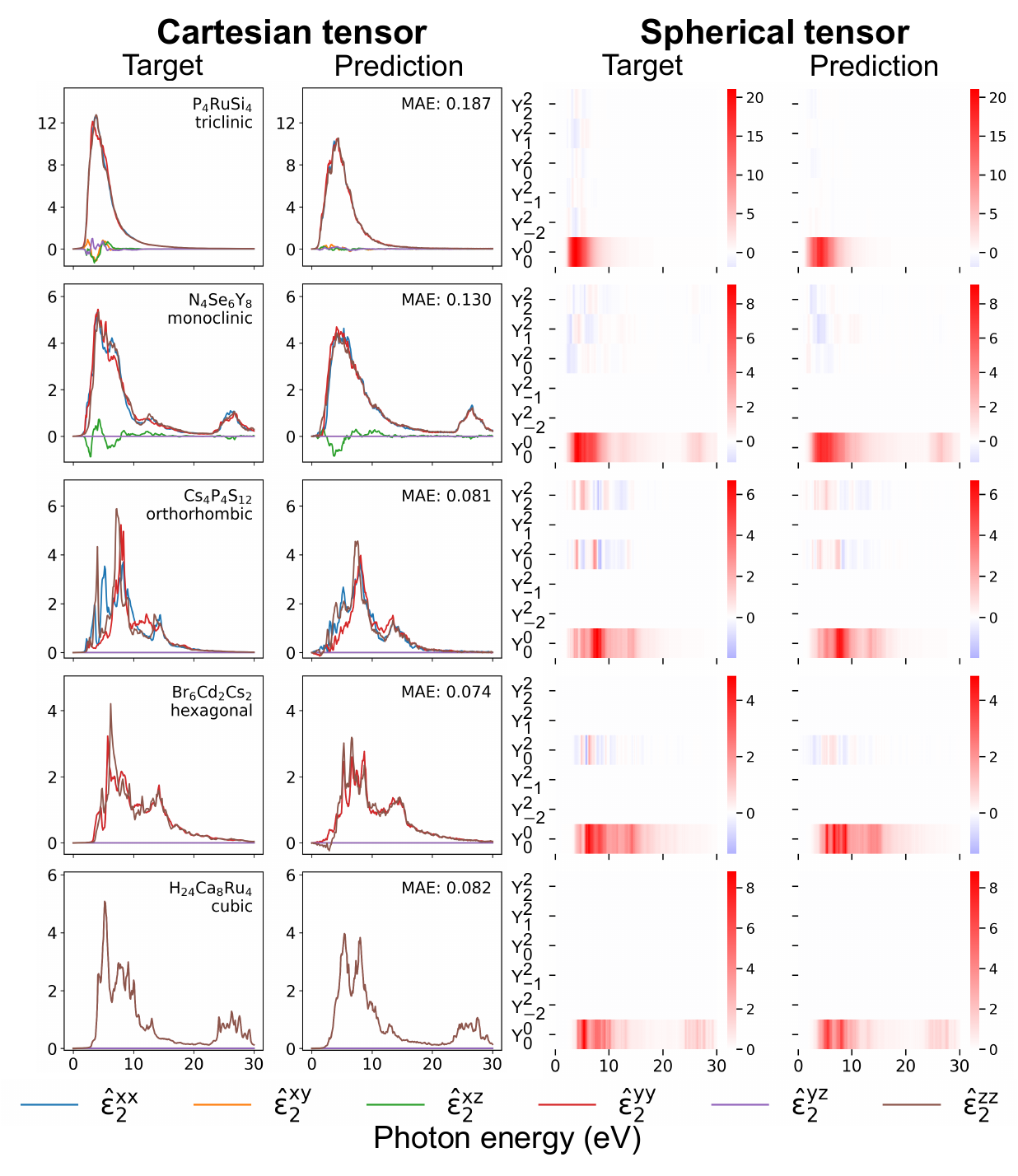}
    \caption{Comparison of predicted and target dielectric tensors in the Cartesian and spherical--harmonics basis in five different crystal systems, ordered by increasing symmetry. In the triclinic system, all tensor components are non-vanishing. The monoclinic system retains four non-vanishing components: \( xx \), \( yy \), \( zz \), and \( xz \). The orthorhombic system exhibits anisotropy with \( xx \neq yy \neq zz \), while the hexagonal system satisfies \( xx = yy \neq zz \). Finally, the cubic system represents the highest symmetry with \( xx = yy = zz \). Left panel shows the pairwise comparison under the spherical tensor representation, while the right panel shows the comparison under the Cartesian tensor. Inset in the right panel provides the chemical formula, crystal system, and MAE.}
    \label{fig:prediction_cart_sph}
\end{figure}

To visualize the model's performance and its ability to preserve symmetry across different crystal systems, we randomly selected five materials from the test set, each representing a distinct crystal class. These materials are arranged according to increasing symmetry, from triclinic (minimum symmetry) to cubic (maximum symmetry), see Fig.~\ref{fig:prediction_cart_sph}. In the Cartesian tensor representation, the predicted dielectric tensor exhibit symmetry-consistent behavior, correctly reproducing the expected nonzero tensor components in agreement with the ground truth.

To further confirm symmetry preservation, we applied a symmetry-based masking procedure to the predicted tensor at each photon energy (see Methods). While equivariance enforces rotational symmetry at the representation level, the network is not explicitly given point-group information. In practice, small numerical deviations or approximate symmetries in the input structures can lead to residual off-diagonal terms. The masking procedure ensures that the predicted spectra conform to the exact crystal symmetry across the entire spectral range. These distinct symmetries are also reflected in the spherical--harmonics decomposition. Each crystal system preserves different spherical channels that depend on its symmetry. This feature is observable in the spherical tensor plots of Fig. \ref{fig:prediction_cart_sph}, where the colormap indicates the magnitude of each spherical--harmonic coefficients, providing a compact, symmetry-adapted view of the underlying Cartesian dielectric tensor. The prominent red region in the \( Y_0^0 \) channel corresponds to a strong isotropic contribution, originating from the trace of the tensor, i.e., \( \hat{\varepsilon}^{xx} + \hat{\varepsilon}^{yy} + \hat{\varepsilon}^{zz} \). In contrast, strong features in the \( Y_0^2 \) channel (red or blue in Fig. \ref{fig:prediction_cart_sph}) indicate significant anisotropic contributions from the diagonal components (e.g. \( -\frac{1}{2}(\hat{\varepsilon}^{xx} + \hat{\varepsilon}^{yy}) + \hat{\varepsilon}^{zz}\)). A detailed breakdown of these contributions can be found in Ref.~\cite{mochizukiSphericalHarmonicDecomposition1988a}. For example, in a hexagonal material, such as \ce{Br6Cd2Cs2}, a visible peak appears in the \( Y_0^2 \) channel, highlighted in red and blue in Fig. \ref{fig:prediction_cart_sph}, which reflects the anisotropic nature of the diagonal components due to the symmetry condition (\( \hat{\varepsilon}^{xx} = \hat{\varepsilon}^{yy} \neq \hat{\varepsilon}^{zz}\)). This illustrates how the spherical--harmonics basis captures not only the symmetry but also the complexity of the directional responses.

As already pointed out, in the spherical tensor representation, the dielectric tensor can be decomposed into \( \ell = 0 \) and \( \ell = 2 \) channels. Since there are no additional irreducible components at this order, the \( \ell = 0 \) channel remains decoupled as it cannot mix with higher-order terms. Improper mixing between the channels could introduce unphysical features in the predicted spectra. However, because both the input crystal structure and the corresponding dielectric tensor are inherently symmetry-consistent, the model preserves the underlying symmetry constraints during training. As a result, it outputs only the physically allowed tensor components, maintaining the correct form of the spherical decomposition. This symmetry-preserving behavior ensures that no spurious mixing occurs between channels and that the reconstructed Cartesian tensor contains only the symmetry-permitted elements. Consequently, the model captures the intrinsic symmetry of materials without violating crystallographic constraints.

\section{Discussion}
\label{sec:discussion}
A major strength of our TSENN model lies in its equivariant architecture, which preserves the symmetries of the input crystal structure. We validate this feature of TSENN through a symmetry-mask based analysis of the predicted dielectric tensors, particularly in the triclinic systems where all components, including the off-diagonal terms, are expected to be nonzero. As a result, the model effectively captures directional couplings (e.g., \(x \rightarrow z\), \(x \rightarrow y\)) that are often missed or underestimated by the graph neural network in the absence of the equivariance architecture. Compared to first-principles methods, TSENN offers an efficient data--driven approach, enabling rapid screening of materials with pronounced anisotropic optical responses. In addition to maintaining a low MAE, the model respects both physical and symmetry constraints. For instance, all predicted spectra decay toward zero as the photon energy approaches 30~eV, a behavior consistent with the asymptotic vanishing of interband transitions at high frequencies. This trend ensures compliance with the \(f\)-sum rule, which dictates that the integral over the optical conductivity (or dielectric function) must satisfy a conserved quantity related to the total number of electrons. Such behavior enables the evaluation of derived quantities governed by the \(f\)-sum rule, including the generalized quantum weights~\cite{PhysRevX.14.011052}. 
For respecting symmetry constraints, the bipartite channel design of our model ensures that the trace (\(\ell=0\)) as well as the traceless tensor part (\(\ell=2\)) remains strictly separated, as validated in SI Sec.~2.3. This separation reflects the irreducible decomposition of a symmetric rank-2 tensor under SO(3), where the scalar \(\ell=0\) channel is invariant under rotations and the \(\ell=2\) channel transforms as a rank-2 tensor. As a result, the model faithfully captures the input symmetry: for example, in hexagonal systems, rotational invariance within the plane perpendicular to the $c$-axis enforces $\hat{\varepsilon}^{xx} = \hat{\varepsilon}^{yy}$, while allowing $\hat{\varepsilon}^{zz}$ to differ. More generally, the model reproduces symmetry-enforced relations across various crystal classes without imposing external symmetry priors.

\begin{figure}[H]
    \centering
    \includegraphics[width=1\linewidth]{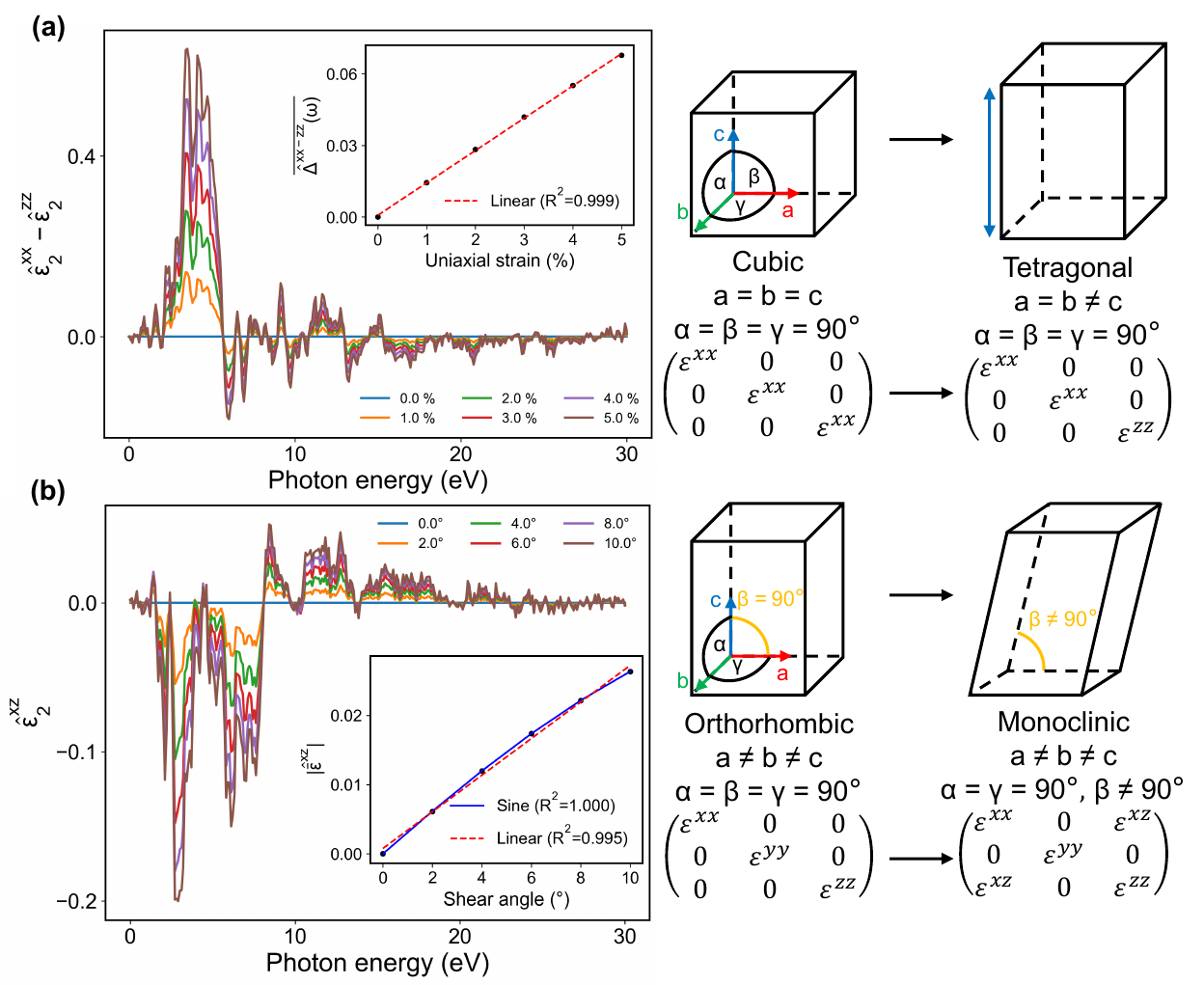}
    \caption{
    Validation of the model's symmetry-preserving behavior under tensile and shear strain. (a) Uniaxial strain along the \(c\)-axis applied to a cubic material which is not included in the training set, varying from 0\% to 5\% in 1\% increments, thereby transforming the structure into a tetragonal system.  At zero strain, cubic symmetry imposes \(\hat{\varepsilon}^{xx} = \hat{\varepsilon}^{yy} = \hat{\varepsilon}^{zz}\), resulting in a vanishing mean anisotropy that grows between \(\hat{\varepsilon}^{zz}\) and \(\hat{\varepsilon}^{xx}\). As strain increases, the mean anisotropy grows linearly with strain, yielding \(R^2 = 0.999\). (b) Shear strain applied to an orthorhombic material by varying the lattice angle $\beta$ from $90^\circ$ to $100^\circ$. 
    In the undistorted case ($\beta=90^\circ$), the off-diagonal dielectric tensor element $\hat{\varepsilon}^{xz}(\omega)$ vanishes. 
    As $\beta$ deviates from $90^\circ$, a finite $\hat{\varepsilon}^{xz}(\omega)$ emerges, and its mean magnitude $|\hat{\overline{\varepsilon}}^{xz}|$ (averaged over the photon energy grid) increases with the shear angle. 
    In the small-strain regime ($<10^\circ$), the dependence is nearly linear ($R^2 = 0.995$), while a sinusoidal fit provides a better fit ($R^2 = 1.000$), consistent with the expected angle dependence. 
    These results confirm that the model captures the onset of symmetry breaking with high fidelity and reproduces the correct physical response across both perturbative and large-angle regimes. 
    }
    \label{fig:shear_test}
\end{figure}

To further investigate the model's treatment of symmetry, we systematically examined the importance of the input structure in satisfying crystal symmetry constraints. Since the model is not explicitly trained on space or point group information, and it is not symmetry-restricted in its architecture, we conducted a series of controlled tests to assess its behavior (Fig.~\ref{fig:shear_test}). Specifically, we selected a cubic material absent from the test set and applied uniaxial strain along the \(c\)-axis, systematically varying the strain from 0\% to 5\% in increments of 1\%, thereby transforming the structure into a tetragonal system, as illustrated in Fig.~\ref{fig:shear_test}(a). From symmetry considerations, the unstrained cubic structure enforces the constraint \(\hat{\varepsilon}^{xx} = \hat{\varepsilon}^{yy} = \hat{\varepsilon}^{zz}\) on the predicted dielectric tensor \(\hat{\varepsilon}^{ij}\), while the strained tetragonal structure satisfies \(\hat{\varepsilon}^{xx} = \hat{\varepsilon}^{yy} \neq \hat{\varepsilon}^{zz}\). At zero strain, the predicted \(\hat{\varepsilon}^{zz}\) and \(\hat{\varepsilon}^{xx}\) components are identical, resulting in a vanishing frequency-averaged anisotropy, $\overline{\hat\Delta^{xx-zz}(\omega)}$, between them, which reflects the preservation of cubic symmetry. As the strain increases, an increasing difference between \(\hat{\varepsilon}^{zz}\) and \(\hat{\varepsilon}^{xx}\) emerges, consistent with the onset of anisotropy expected for a tetragonal system. By computing the frequency-averaged anisotropy components across all strained structures, we observe a highly linear dependence on the strain magnitude, with a \(R^2 = 0.999\). We also performed additional \emph{ab initio} calculations to verify that the predicted trends are physically consistent (see SI, Sec. 5). These results demonstrate that the model faithfully captures the progressive symmetry breaking induced by strain while preserving the underlying physical constraints.

We also evaluated the model's response to shear strain perturbations by modifying the lattice angles, as shown in Fig.~\ref{fig:shear_test}(b). Starting from an orthorhombic material (\(a \neq b \neq c\), \(\alpha = \beta = \gamma = 90^\circ\)), we introduced a monoclinic distortion by incrementally varying \(\beta\) from \(0^\circ\) to \(10^\circ\) in \(2^\circ\) steps. The essential distinction between the orthorhombic and monoclinic systems is that in the latter, \(\beta\) deviates from \(90^\circ\), allowing the emergence of a nonzero \(\hat{\varepsilon}^{xz}(\omega)\) component in the dielectric tensor. When no perturbation is applied (\(\beta = 90^\circ\)), \(\hat{\varepsilon}^{xz}(\omega)\) vanishes, yielding a zero mean magnitude \(| \hat{\overline{\varepsilon}}^{xz} |\). As the angular deviation increases, a finite \(\hat{\varepsilon}^{xz}(\omega)\) progressively emerges, producing a nonzero \(|\hat{\overline{\varepsilon}}^{xz}|\). We quantified this emergent behavior as a function of the shear angle and observed an almost perfect linear relationship in the small-strain regime (\(R^2 = 0.995\)). As expected, the emergence of the off-diagonal component is directly coupled to the applied \(\beta\)-angle distortion; consequently, a sine fit provides an even better description, with \(R^2 = 1.0\). At larger distortions, the dependence smoothly evolves toward a sinusoidal form, reflecting higher-order effects of the angular perturbation. \emph{Ab initio} calculations confirm this crossover behavior: linear scaling in the perturbative regime and sinusoidal dependence at larger angles (see SI Sec.~5). The model thus reproduces both regimes with high fidelity, demonstrating that it not only respects the symmetry of the input structure but also quantitatively matches the physical response to both small and large shear distortions.

While the primary focus of this work is on predicting the imaginary part of the dielectric tensor, we also demonstrate that the real part can be accurately reconstructed by using K-K relations. This yields a complete complex dielectric tensor, which in turn enables the derivation of a broad range of optical response functions containing rich information, such as the direct extraction of optical bandgaps (see SI Sec.~5). This predictive capability is valuable for designing materials for photovoltaics, sensors, and nonlinear optical applications. Triclinic crystals, for example, with both \(\hat{\varepsilon}^{xz}\) and \(\hat{\varepsilon}^{xy}\) components, offer multidirectional optical control, which is desirable for heterostructures and chiral photonic devices. 

An important challenge revealed by our analysis is that predictive errors nearly double in the low-frequency regime ($\omega < \omega_p$), where the dielectric response is dominated by rich, material-specific absorption features with strong oscillator strengths. Here, $\omega_p$ denotes the plasma frequency, which marks the crossover between collective electronic oscillations and the asymptotic high-frequency regime. The low frequency regime is both physically meaningful--since it encodes the distinct fingerprints of each compound--and difficult to capture accurately. One might consider introducing an energy-resolved weighting scheme during training to place more emphasis on this region; however, such an approach would require prior knowledge of $\omega_p$ for each material and would introduce material-dependent biases at inference time, thereby compromising generalizability. To prevent this, we chose to retain an energy-independent training loss and instead provide a detailed frequency-resolved benchmarking in SI Sec.~4, which compares the error behavior above and below the plasma frequency.

It is appropriate to acknowledge the intrinsic limitations of the underlying \emph{ab initio} data, which is based on the IPA rather than the RPA, so that the training of our model neglects local-field effects. Although the RPA provides some improvement by narrowing the discrepancy between theory and experiment, the computationally far more demanding Bethe--Salpeter approach~\cite{onidaElectronicExcitationsDensityfunctional2002} is needed to capture excitonic effects properly. The Kohn--Sham DFT-based eigenvalues systematically underestimate bandgaps that can be corrected through more expensive $GW$ calculations. Despite these limitations, TSENN-trained on a relatively modest dataset of approximately \(10^3\) materials achieves strong data efficiency in learning high-rank tensorial responses. Recent work shows that machine-learning models trained on IPA spectra can be fine-tuned or transferred to RPA-level data to incorporate local-field effects~\cite{grunertMachineLearningClimbs2025}. In addition to these physical approximations, the dataset itself is dominated by materials with weak optical anisotropy, leading to intrinsically small off-diagonal components of the dielectric tensor and an imbalance in the distribution of tensor magnitudes. To more accurately assess model performance on genuinely anisotropic systems, we curated an auxiliary dataset composed exclusively of materials with finite off-diagonal responses. This benchmark not only targets applications that depend on off-diagonal tensor elements, but also reveals that our model attains substantially improved accuracy on these components (see SI Sec.~8). Building on this enhanced capability to predict off-diagonal optical responses, we further demonstrate the potential applicability of our framework by considering the magneto-optical Kerr effect and magnetic circular dichroism; see SI Sec.~9.

In summary, TSENN advances the prediction of complex frequency-dependent tensorial optical properties by accurately resolving full tensor components across various symmetry classes, providing a practical tool for accelerating materials discovery for next-generation optoelectronic technologies. Our approach is inherently general as it proceeds by decomposing Cartesian tensors into spherical--harmonics components, and it would thus be naturally extendable to treat higher-rank tensors such as the piezoelectric and elastic tensors through the inclusion of higher-order $\ell$ channels. Furthermore, it holds promise for modeling sequential higher-order properties such as the shift current~\cite{fangDatasetTensorialOptical2025} and second harmonic generation (see SI Sec.~10.), making it a versatile framework for learning symmetry-constrained anisotropic physical phenomena.

\section{Methods}\label{sec3}
\subsection{Ab initio Calculations}
DFT calculations were performed using OpenMX \cite{openmxfirst, openmxsecond, openmxthird, openmxfourth}, which employs numerical pseudo-atomic orbitals as the basis set to expand the one-particle Kohn-Sham wave functions. The high-throughput calculations involved three main steps: generating input files, submitting jobs, and monitoring their progress. The input files were generated using the \texttt{ASE} package, which provides a predefined lookup table for optimized basis selection. However, this table has not been updated since 2013, leading to missing entries for some elements with heavy \( f \)-orbital character. We used the PBE-GGA functional in all calculations. We use a custom workflow for job submission and monitoring, implemented with \texttt{Atomate}~\cite{MATHEW2017140}. Self-consistent field (SCF) convergence was ensured to an accuracy of \( 3.67 \times 10^{-8} \) Hartree or \(10^{-6}\) eV. All other settings used the default OpenMX parameters. The SCF \(k\)-point grid was generated using the \texttt{pymatgen} automatic density function with a target density of 2000 \(k\)-points per atom (kppa). This method selects \(\Gamma\)-centered meshes for hexagonal and face-centered lattices, and Monkhorst--Pack grids for all other crystal systems, ensuring consistent and resolution-adaptive Brillouin zone sampling.

After achieving SCF convergence with OpenMX, we extracted the Hamiltonian, overlap matrices, and position operators to evaluate the tensorial relative permittivity. The dielectric tensor was computed within the IPA using the Kubo formula: 
\begin{equation}
\varepsilon_{\mathrm{r}}^{\alpha\beta}(\omega)
= \frac{\varepsilon^{\alpha\beta}(\omega)}{\varepsilon_0}
= \chi^{\alpha\beta}(\omega) + \delta^{\alpha\beta},
\qquad
\chi^{\alpha\beta}(\omega)
= \frac{e^2}{\varepsilon_0 \hbar}
\int \frac{d^3\mathbf{k}}{(2\pi)^3}
\sum_{n,m}
\frac{f_{nm}\, r_{nm}^\alpha r_{mn}^\beta}{\omega_{mn}-\omega - i\eta}\, .
\end{equation}

In this work, we denote the real and imaginary parts of the $\varepsilon^{\alpha\beta }_{\text{r}}$
as $\varepsilon_1^{\alpha\beta}$ and $\varepsilon_2^{\alpha\beta}$, respectively. Here \(\hbar \omega_{nm} = E_n - E_m\) denotes the interband energy difference, 
\(f_{nm} = f(E_n) - f(E_m)\) is the difference of Fermi--Dirac occupations, and 
\(r^\alpha_{nm}\) is the Berry connection, defined as
\(r^\alpha_{nm} = i \left\langle u_{n \mathbf{k}} \middle| \partial_\alpha u_{m \mathbf{k}} \right\rangle,\)
with \(\partial_\alpha = \partial / \partial k_\alpha\) the derivative with respect to the Cartesian momentum coordinate \(\alpha \in \{x, y, z\}\),  where the Hamiltonian and position operator are projected onto a localized basis and the Brillouin-zone integrals are evaluated efficiently 
\cite{wangFirstprinciplesCalculationOptical2019, wangFirstprinciplesCalculationNonlinear2017, sipeNonlinearOpticalResponse1993, marzariMaximallyLocalizedGeneralized1997, marzariMaximallyLocalizedWannier2012}. A uniform \(k\)-point mesh was constructed with reciprocal-space resolution of \(0.05~\text{\AA}^{-1}\) along each reciprocal lattice vector, ensuring consistent Brillouin-zone sampling across all materials. The smearing parameter \(\eta\), introduced to improve numerical convergence of the optical response, was fixed at 0.05~eV in all calculations. This value  corresponds to typical carrier relaxation times ($\sim$10~femtoseconds) in semiconductors \cite{serneliusIntrabandRelaxationTime1991,sivadasGateControllableMagnetoopticKerr2016, zhouHighthroughputPredictionCarrier2020}. This procedure is equivalent to constructing a Wannier-based tight-binding representation of the Hamiltonian, which provides a localized framework for efficient evaluation of the dielectric response. The present calculations are performed within the IPA, i.e. based on Kohn--Sham eigenvalues and wavefunctions without including quasiparticle (GW) corrections or electron--hole interaction (excitonic) effects \cite{agaFirstPrinciplesStudyQuasiParticle2023}. Consequently, discrepancies between the computed dielectric spectra and the corresponding experimental results are expected.

\subsection{Machine Learning}\label{sec:ML}
The spectra were computed with a resolution of 0.01 eV, with a total of 3,001 points. Including all points in the training process would significantly increase computational cost. To make training more efficient, we performed downsampling by interpolating the data along the photon energy grid to 300 points, which still captures the spectral response reasonably. We used an 80:10:10 split for training, validation, and testing datasets. Given our focus on crystal symmetries, we implemented balanced loading by analyzing the distribution of crystal systems across all splits to ensure even representation, see SI Sec. 7 for details. 

Group theoretical analysis reveals five distinct cases based on crystalline symmetries. This enables us to construct a mask $\mathbf{M}$ that identifies the symmetry-allowed, non-vanishing components of the tensorial spectra. The mask is derived by initializing a \(3 \times 3\) matrix of ones, \(\mathbf{1}_{3 \times 3}\), and applying the set of rotation operations \(\mathcal{O} = \{ R_1, R_2, \dots, R_{N_R} \}\) associated with the space group symmetry of the crystal. The final mask is computed by averaging over the transformed matrices:
\[
\mathbf{M} = \frac{1}{N_R} \sum_{R_i \in \mathcal{O}} R_i^\top \mathbf{1}_{3 \times 3} R_i.
\]
For example, in a monoclinic system, this process yields the following mask:
\begin{equation}
\mathbf{M} = \begin{pmatrix}
1 & 0 & 1 \\
0 & 1 & 0 \\
1 & 0 & 1
\end{pmatrix}
\end{equation}
This mask highlights the non-vanishing tensor components, such as \(xx\), \(yy\), \(zz\), and \(xz\), as dictated by the symmetry constraints. To identify any symmetry violations, we define the residual mask as  
\[
\text{Residual mask} = \epsilon(\mathbf{1}_{3\times3} - \mathbf{M}) = \begin{pmatrix}
0 & \epsilon & 0 \\
\epsilon & 0 & \epsilon \\
0 & \epsilon & 0
\end{pmatrix},
\]
where \(\epsilon = 10^{-6} \) serves as the detection threshold. We apply this residual mask across the photon energy grid to verify the input tensors before training and the predicted tensors after training. Our results confirm that all tensors respect the imposed symmetry constraints, demonstrating that the model maintains equivariance throughout training without symmetry breaking.

We performed a comprehensive hyperparameter search using Optuna \cite{akiba2019optunanextgenerationhyperparameteroptimization} to optimize performance of the model. The multiplicity of irreducible representations \(m\) used in the convolution layers was varied from 16 to 64, with 32 selected as the optimal value. The number of pointwise convolution layers \(L\) was tuned between 1 and 4, with the final model using 4. The embedding feature vector length was explored from 16 to 160, with 128 yielding the best performance. Dropout probability was tested at 0, 0.1, 0.2, and 0.4 with 0.4 selected as the optimal choice. Batch sizes from 1 to 16 were systematically evaluated, and a size of 8 was selected as the best tradeoff between computational efficiency and training stability. Smaller batches (e.g., size 1) led to unstable optimization due to insufficient statistical averaging per update. For very large batches ($>32$), the anisotropy was found to become less pronounced compared to smaller ones, likely because the optimization tends to get trapped in local minima and averages out sharper anisotropic signals through stochastic smoothing. The hyperparameters of the AdamW optimizer were carefully tuned. Learning rates between \(5 \times 10^{-4}\) and \(5 \times 10^{-2}\) were explored, with \(1 \times 10^{-2}\) achieving the lowest validation loss. To further enhance convergence, we applied PyTorch's \texttt{CosineAnnealingWarmRestarts} scheduler, which gradually reduces the learning rate according to a cosine decay schedule. The initial number of iterations before the first restart was set to \(T_0 = 10\), with subsequent settings kept at their default values. This smooth, non-monotonic decay enables larger updates during the early training stages and finer adjustments at later stages, helping the model escape shallow minima and converge more robustly~\cite{loshchilov2017sgdrstochasticgradientdescent}.

To contrast the computational efficiency, we benchmarked our workflow against a representative \emph{ab initio} calculation. For this purpose, we recalculated the dielectric tensor of \ce{Ag3AsS4} (Materials Project ID: mp-9538, 16 atoms per unit cell). The DFT self-consistent calculation converged in 119~s using a hybrid MPI/OpenMP setup on 2 AMD EPYC~7763 CPUs (256 threads in total) with default workflow settings. The subsequent dielectric tensor evaluation, carried out with Brillouin-zone integration parallelized over 128 processes, required 22~min~9~s. Including workflow orchestration overhead, the full pipeline--from job launch to data storage--took 26~min~32~s. By contrast, the GNN-based workflow was orders of magnitude faster: graph construction required \(6.96 \pm 0.01\)~ms, forward inference \(695.20 \pm 0.01\)~ms, and K-K relations took \(570.04 \pm 0.01\)~ms, giving a total runtime of 1.18~s on a single NVIDIA RTX~4090 GPU. In other words, our approach reduces a half-hour DFT calculation to just over one second.

\section*{Data Availability}
All datasets and optimized model weights used in this study are openly available at \url{https://github.com/qmatyanlab/TSENN}. The repository also includes reproducible Jupyter notebooks demonstrating data usage and reproducing the figures presented in this work.

\section*{Code Availability}
The third-party code \textsc{OpenMX} used in this study is publicly available at \url{https://www.openmx-square.org}. The custom workflows developed to compute the dielectric functions are accessible at \url{https://github.com/qmatyanlab/OpenMX-workflow} and were implemented using the \textsc{Julia} programming language (\url{https://julialang.org}). 

\backmatter

\section*{Declarations}
\bmhead{Acknowledgements}
We are grateful to Alexander J. Heilman for insightful discussions on equivariant transformations. We also thank Weiyi Gong for helpful conversations on training strategies and for sharing expertise in equivariant graph neural networks. 

\bmhead{Funding}
T.-W. H., A.B., and Q.Y. acknowledge support from the U.S. National Science Foundation under the Expanding Capacity in Quantum Information Science and Engineering (ExpandQISE) program (Award No. NSF-OMA-2329067). Z.F. acknowledges support from the U.S. Department of Energy, Office of Science, Basic Energy Sciences, under Award No. DE-SC0023664. This research utilized resources provided by the National Energy Research Scientific Computing Center (NERSC), a U.S. Department of Energy Office of Science User Facility operated under Contract No. DE-AC02-05CH11231, through NERSC Award No. BES-ERCAP0029544. Additional computational support was provided by the Northeastern University Discovery Cluster. This work benefited from the resources provided by the Massachusetts Technology Collaborative (Award No. MTC-22032) and Northeastern University’s Quantum Materials and Sensing Institute (QMSI).

\bmhead{Conflict of Interest}
The authors declare no conflict of interest.

\bmhead{Author contribution}
T.-W. H., Z. F., and Q. Y. conceived the study. T.-W. H. performed the calculations and trained the models. Z. F. assisted with data analysis. A. B. and Q. Y. supervised the project. The manuscript was written with contributions from all authors, and all authors approved the final version.


\end{document}


\title[Article Title]{\textbf{Supplementary Information for Accurate Prediction of Tensorial Spectra Using Equivariant Graph Neural Network}}


\author*[1,2]{\fnm{Ting-Wei} \sur{Hsu}}\email{hsu.ting@northeastern.edu}

\author[1,2]{\fnm{Zhenyao} \sur{Fang}}\email{z.fang@northeastern.edu}

\author[1,2]{\fnm{Arun} \sur{Bansil}}\email{ar.bansil@northeastern.edu}

\author*[1,2]{and \fnm{Qimin} \sur{Yan}}\email{q.yan@northeastern.edu}


\affil[1]{\orgdiv{Department of Physics}, \orgname{Northeastern University}, \city{Boston}, \postcode{02155}, \state{Massachusetts}, \country{USA}}

\affil[2]{\orgdiv{Quantum Materials and Sensing Institute}, \orgname{Northeastern University}, \city{Burlington}, \postcode{01803}, \state{Massachusetts}, \country{USA}}

\maketitle
\tableofcontents

\newpage
\section{Details of Evaluation Metrics} \label{SI:sec1}
 
The first metric is the direct mean absolute error (MAE) of the dielectric tensor, together with its per-component counterpart,  
\begin{equation}
\text{MAE}=\frac{1}{6 N_\omega} \sum_\omega \sum_{\alpha \leq \beta}
\left|\hat{\varepsilon}^{\alpha \beta}(\omega)-\varepsilon^{\alpha \beta}(\omega)\right|,
\quad 
\text{MAE}^{\alpha \beta}=\frac{1}{N_\omega} \sum_\omega
\left|\hat{\varepsilon}^{\alpha \beta}(\omega)-\varepsilon^{\alpha \beta}(\omega)\right|.
\end{equation}
Because the magnitudes of the diagonal and off-diagonal tensor components differ significantly, we additionally report a normalized error,  
\begin{equation}
\text{NMAE}^{\alpha\beta} (\%) = 100 \times \frac{\text{MAE}^{\alpha\beta}}{M^{\alpha\beta}},
\end{equation}
where the normalization factor \(M^{\alpha\beta}\) is defined as  
\begin{equation}
M^{\alpha\beta} =  \operatorname{median}_i \Big[ \max_\omega \, 
\varepsilon_{i}^{\alpha\beta}(\omega) \Big].
\end{equation}
That is, for each material index \(i\), we first take the maximum of 
\(\varepsilon_{i}^{\alpha\beta}(\omega)\) across the frequency domain \(\omega\).  
We then compute the median of these maxima over all samples in the dataset. The resulting value \(M^{\alpha\beta}\) serves as a characteristic magnitude, providing a robust measure of the typical spectral amplitude against which errors can be meaningfully normalized.

To isolate anisotropic behavior, we subtract the isotropic contribution from the tensor,  
\begin{equation}
\varepsilon_{\text{aniso}}(\omega)=\varepsilon(\omega)-\frac{1}{3}\operatorname{Tr}[\varepsilon(\omega)] I.
\end{equation}
This yields the component-wise anisotropic contributions $\varepsilon_{\text{aniso}}^{\alpha \beta}(\omega)$, the component-averaged value $\mathrm{MAE}_{\text {aniso }}$, as well as their overall magnitude $\left\|\varepsilon_{\text{aniso}}(\omega)\right\|_F$. Based on this decomposition, we define the anisotropic per-component error as  
\begin{equation}
\text{MAE}_{\text{aniso}}^{\alpha \beta}=\frac{1}{N_\omega} \sum_\omega\left|\hat{\varepsilon}_{\text{aniso}}^{\alpha \beta}(\omega)-\varepsilon_{\text{aniso}}^{\alpha \beta}(\omega)\right|.
\end{equation}
Notably, this measure corresponds to the performance of the $\ell=2$ spherical--harmonic channels, but expressed in the Cartesian basis. To quantify the strength of anisotropy independent of basis choice, we compute the Frobenius norm of the traceless part of the dielectric tensor and evaluate the mean absolute error between the predicted and reference anisotropy magnitudes,  
\begin{equation}
\text{MAE}_{\text {aniso}}^{\text{norm}}=\frac{1}{N_\omega} \sum_\omega\left|\left\|\hat\varepsilon_{\text {aniso}}(\omega)\right\|_F-\left\|\varepsilon_{\text {aniso}}(\omega)\right\|_F\right|.
\end{equation}
Unlike the per-component anisotropic errors, this scalar quantity is rotation invariant, directly assessing the accuracy of the $\ell=2$ channel magnitude. The effectiveness of this metric is demonstrated in Supplementary Fig.~\ref{SI_Plot_3}. In addition to absolute magnitudes, it is also important to evaluate the spectral shape of the predicted response. To this end, we compute the first-derivative MAE of each tensor component ($\text{MAE}^{\prime, \alpha \beta}$),  
\begin{equation}
\text{MAE}^{\prime, \alpha \beta}=\frac{1}{N_\omega-1} \sum_\omega\left|\Delta\hat{\varepsilon}^{\alpha \beta}(\omega)-\Delta\varepsilon^{\alpha \beta}(\omega)\right|,
\end{equation}
where $\Delta\varepsilon^{\alpha \beta}(\omega)=\varepsilon^{\alpha \beta}(\omega+\delta \omega)-\varepsilon^{\alpha \beta}(\omega)$. This metric emphasizes differences in slope rather than magnitude, ensuring that fine spectral features, such as peak positions and line shapes, are faithfully reproduced. Finally, to quantify the similarity in spectral distribution from an information-theoretic perspective, we compute the Kullback-Leibler (KL) divergence, 
\begin{equation}
D_{\mathrm{KL}}(p \parallel q)=\sum_\omega p(\omega) \log \frac{p(\omega)}{q(\omega)},
\end{equation}
where $p(\omega)$ and $q(\omega)$ denote the normalized target and predicted spectra, respectively. Together, these complementary metrics evaluate not only the overall prediction accuracy but also the anisotropy strength, spectral shape, and distributional fidelity.

\begin{figure}[H]
    \centering
    \captionsetup{labelformat=empty}
    \includegraphics[width=1.0\linewidth]{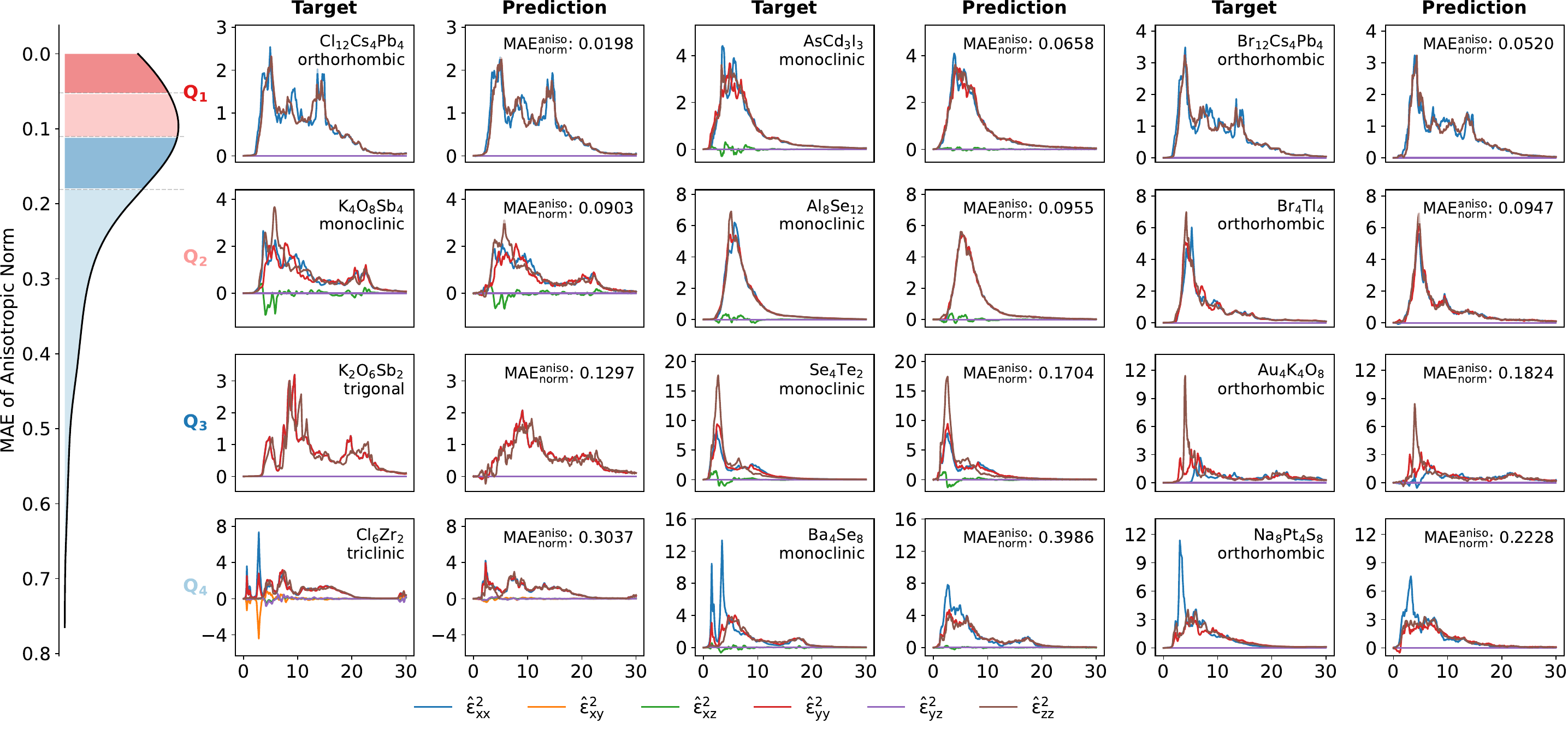}
    \caption{\textbf{Supplementary Figure 1.}
    Representative dielectric spectra selected based on the anisotropy--norm--error $\text{MAE}^{\text{norm}}_{\text{aniso}}$. 
    Targets (left) display the physically correct anisotropy, while model predictions (right) are benchmarked under this scalar measure. 
    Sampling by this metric provides an interpretable view of how well anisotropy strength is captured across frequency and complements the component-wise MAE reported in the main text.}
    \label{SI_Plot_1}
\end{figure}

Supplementary Fig.~\ref{SI_Plot_1} illustrates representative predictions sampled using $\text{MAE}^{\text{norm}}_{\text{aniso}}$ as the selection metric. This approach allows us to construct cumulative kernel-density-estimator (KDE) plot to identify what constitutes a "good" fit under this anisotropy-focused measure, while also providing additional opportunities to visualize prediction results beyond the Cartesian-component view.

\color{black}

Using the last complementary measure of the model's performance, we compress the sequential tensorial outputs into single scalar quantities for both the predicted and target spectra. Specifically, for each sample, we first average all tensor components over the photon energy range and then compute the mean over the tensor indices to obtain a single scalar value, denoted as $\bar{\bar{\varepsilon}}^{(k)}$ with $k \in \{\text{prediction}, \text{target}\}$. The scalar quantity is computed as
\begin{equation}
    \bar{\varepsilon}^{(k)} = \frac{1}{6\,n_\omega} \sum_{i \leq j} \sum_{\omega} \varepsilon_{ij}^{(k)}(\omega),
\end{equation}
where $n_\omega$ is the number of sampled photon energies and the factor of six accounts for the unique tensor components.

Following this compression, we evaluate the model's performance using the mean absolute error (MAE) in physical units, defined as
\begin{equation}
    \text{MAE}_{\text{scalar}} = \left| \bar{\varepsilon}^{\text{prediction}} - \bar{\varepsilon}^{\text{target}} \right|.
\end{equation}
We also compute the relative error,
\begin{equation}
    \text{Relative Error} = 
    \frac{\left| \bar{\bar{\varepsilon}}^{\text{prediction}} - \bar{\bar{\varepsilon}}^{\text{target}} \right|}
         {\left| \bar{\bar{\varepsilon}}^{\text{target}} \right|},
\end{equation}
which provides a normalized measure of the prediction accuracy relative to the overall magnitude of the true tensor response.

This scalar-level analysis enables a concise visualization of performance, such as scatter plots comparing predicted and target $\bar{\bar{\varepsilon}}$ values, and cumulative distributions of relative errors (see Supplementary Fig.~\ref{SI_Plot_2}). A nuance associated with this of is that both the directional and frequency dependence are condensed into a single scalar quantity, resulting in the loss of information regarding anisotropy and off--diagonal tensor components. Consequently, this metric should be regarded solely as a complementary measure, reflecting the model's overall isotropic prediction accuracy.

\begin{figure}[H]
    \centering
    \captionsetup{labelformat=empty}
    \includegraphics[width=0.8\linewidth]{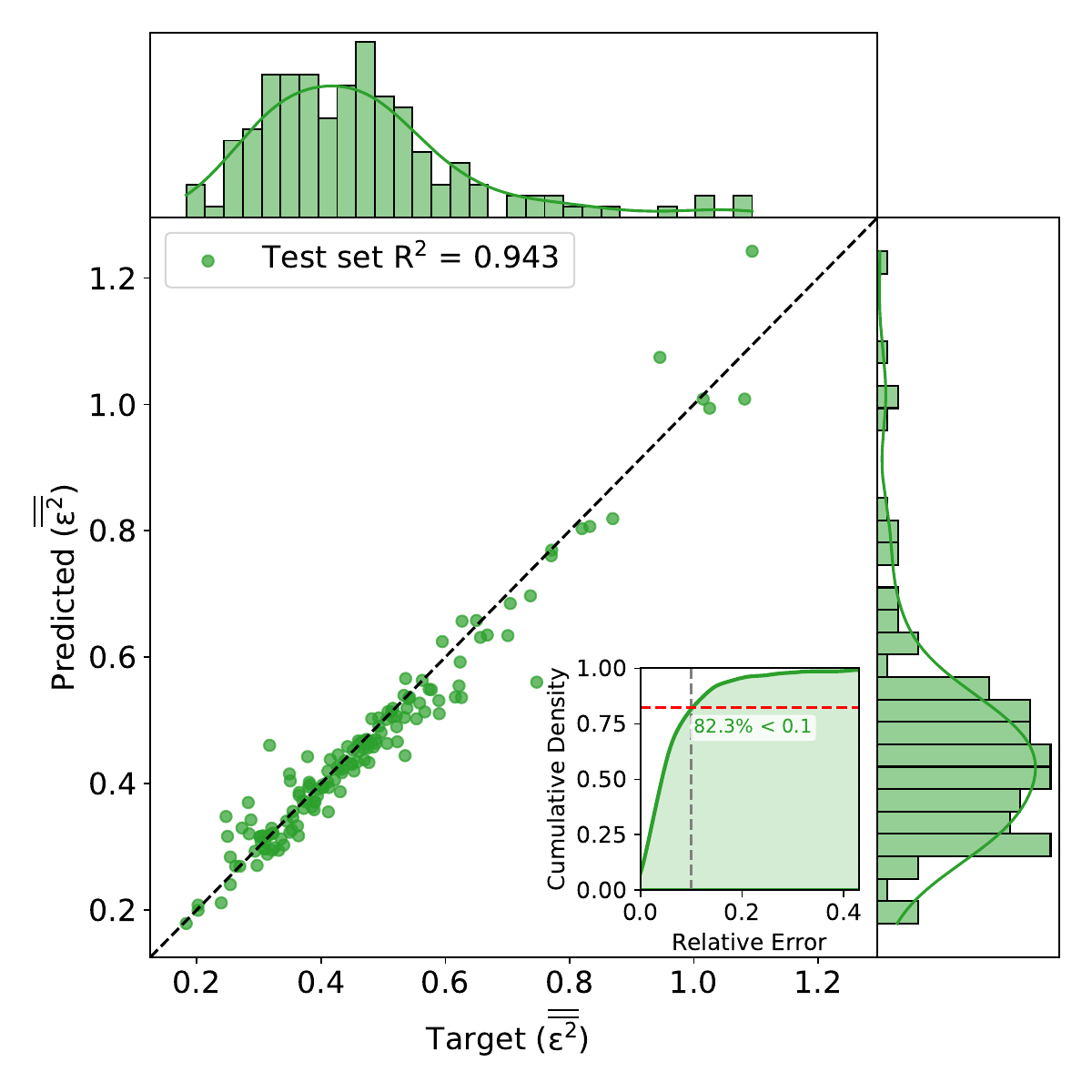}
    \caption{\textbf{Supplementary Figure 2}
    Scatter plot comparing the target and predicted values of $\bar{\bar{\varepsilon}}$, showing an $R^2$ score of $0.943$. The dashed line represents the ideal $y=x$ relationship. Marginal histograms illustrate the distributions of target values (top) and predicted values (right). The inset shows the cumulative distribution of relative errors between the target and predicted $\bar{\bar{\varepsilon}}$, indicating that $82.3\%$ of the data fall within a relative error threshold of $10\%$ (red dashed line).}
    \label{SI_Plot_2}
\end{figure}

\color{black}

\section{Ablation Study}\label{ablation_study}
\subsection{Performance Comparison Between Equivariant and Scalar Models}
In this ablation study, we examine how the choice of output representation, the inclusion of a loss balancer, and the selection of the maximal irreducible representation \(\ell_{\text{max}}\) influence model performance.

We compared two variants of our model: the equivariant version, TSENN, and a scalar-output baseline, TSENN--Scalar (TSENN--S), which shares the same architecture as GNNOpt. TSENN outputs both isotropic and anisotropic components using \(N_\omega \times 0e + N_\omega \times 2e\), where \(N_\omega\) is the number of photon energy points. In contrast, TSENN--S produces a sequence of scalars using \(6N_\omega \times 0e\), entirely removing equivariance and relying solely on the model to infer directional dependencies from data. As illustrated in Supplementary Fig.~\ref{SI_Plot_3}, TSENN--S captures the continuous spectral profile well but fails to reproduce the underlying tensorial structure. While both models achieve comparable accuracy in reconstructing the overall spectral shape—as monitored by \(\text{MAE}\)--this alone is insufficient to assess their ability to capture anisotropy and an additional metric is required to evaluate performance on the anisotropic component.

\begin{figure}[H]
    \centering
    \captionsetup{labelformat=empty}
    \includegraphics[width=0.8\linewidth]{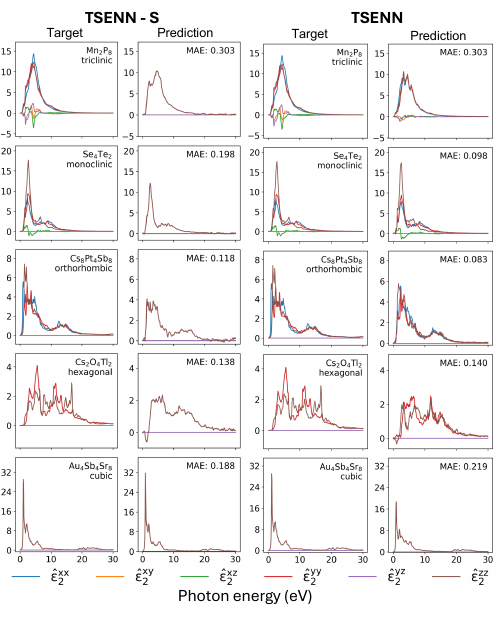}
    \caption{\textbf{Supplementary Figure 3.} Performance comparison between the equivariant model (TSENN) and the invariant baseline (TSENN--S) on five test crystals. TSENN--S captures the smooth spectral envelope but fails to preserve tensorial anisotropy, yielding isotropic predictions across all cases.}
    \label{SI_Plot_3}
\end{figure}

We adopt the previously defined $\text{MAE}^{\text{norm}}_{\text{aniso}}$ from SI Section~\ref{SI:sec1}. 
By plotting the anisotropy strength as a function of photon energy, we can directly compare the agreement between the target spectra and model predictions. 
As shown in Supplementary Fig.~\ref{SI_Plot_4}(a), faithful reproduction of the anisotropy strength is sufficient to reconstruct the full dielectric tensor spectra with high fidelity. 
Panel (b) further illustrates this by comparing individual tensor components, where close agreement between the target and prediction confirms the reliability of $\text{MAE}^{\text{norm}}_{\text{aniso}}$ as a compact yet informative diagnostic for evaluating anisotropic features.

\begin{figure}[H]
    \centering
    \captionsetup{labelformat=empty}
    \includegraphics[width=0.8\linewidth]{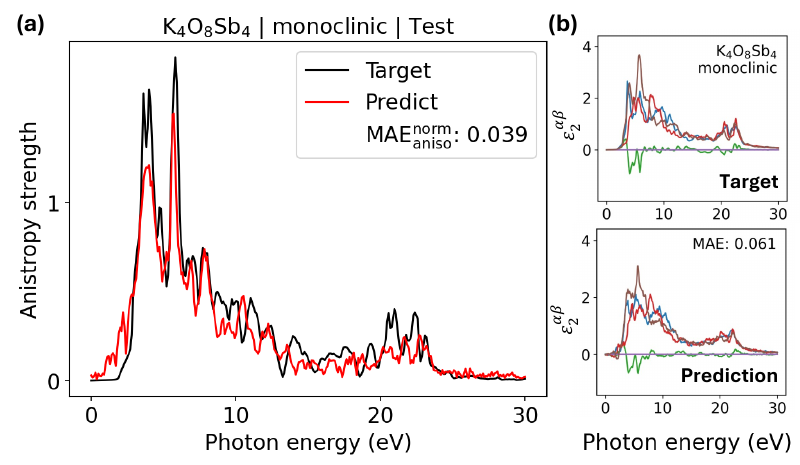}
    \caption{ \textbf{Supplementary Figure 4.} (a) Comparison of the target (black) and predicted (red) anisotropy strength for a representative monoclinic compound \ce{K4O8Sb4} in the test set. The close agreement indicates that recovering the anisotropy strength is sufficient to reproduce the full tensor spectra with high fidelity. (b) The off--diagonal $xz$ component is nearly identical to the target, and the diagonal elements satisfy $xx \neq yy \neq zz$, consistent with the expected monoclinic symmetry. The color coding follows the same convention as in the other plots throughout this work.}
    \label{SI_Plot_4}
\end{figure}

We examine two representative cases from the test set in Supplementary Fig.~\ref{SI_Plot_5}. The first is a cubic material, \ce{Li3NbS4} (Materials Project ID: mp-755309), which is expected to be nearly isotropic. Here, both TSENN and TSENN--S correctly predict the isotropic component, but TSENN--S outputs anisotropic strengths close to zero, with minor oscillations resembling noise. In contrast, the second case, \ce{K4O8Sb4} (mp-10417), is a monoclinic crystal with significant anisotropy. Now, TSENN--S again predicts near-zero anisotropic strength across all frequencies, failing to capture any directional dependence. TSENN, however, successfully captures the frequency-dependent anisotropic response, although with some magnitude deviation. These results demonstrate that choosing a scalar-only output breaks the equivariant symmetry of the message-passing process and limits the model's ability to predict tensorial behavior. Overall, anisotropic strength proves to be an effective indicator for evaluating how well the model captures anisotropic characteristics of the spectra.

\begin{figure}[h]
    \centering
    \captionsetup{labelformat=empty}
    \includegraphics[width=0.8\linewidth]{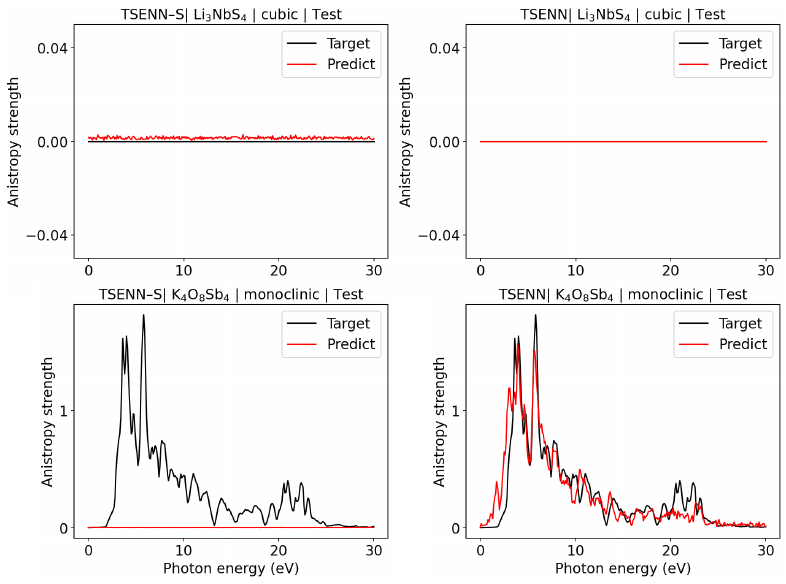}
    \caption{\textbf{Supplementary Figure 5.} Comparison between TSENN and TSENN--S on two representative crystal systems from the test set. The top panels show results for \ce{Li3NbS4} (mp-755309), a cubic material with fully isotropic optical response. As expected, the anisotropic strength should vanish across all photon energies. TSENN produces near-zero anisotropic strength with stable predictions, while TSENN--S yields small, non-zero noise around zero, indicating its inability to preserve isotropy cleanly. The bottom panels show results for \ce{K4O8Sb4} (mp-10417), a monoclinic material with strong anisotropy. TSENN successfully captures the frequency-dependent anisotropic behavior, whereas TSENN--S fails to recover any directional information, predicting nearly zero anisotropic strength across all energies. This comparison highlights the critical role of equivariant output design in capturing tensorial features.}
    \label{SI_Plot_5}
\end{figure}

Supplementary Table~\ref{tab1} compares the spectral-prediction performance of TSENN and TSENN--S using the following evaluation metrics. (i) Following the definition in SI section~\ref{SI:sec1}, we compute the MAE between the predicted and target dielectric tensors, which quantifies the overall spectral accuracy. (ii) To assess the model's performance on the anisotropic component, we use the $\text {MAE}^{\text {norm}}_{\text{aniso }}$. And (iii) to compute the cumulative density \(\rho_{\text{aniso}}^{\text{norm}}\), we found that using relative errors introduced numerical instability due to division by values near zero. Instead, we adopt an alternative threshold-based approach:
\begin{equation}
\rho_{\text{aniso}}^{\text{norm}} = \left( \frac{1}{N} \sum_{N}\mathbb{I}\left( \operatorname{MAE}_{\text{aniso}}^{\text{norm}} < \tau \right) \right) \times 100,
\end{equation}
where \(\mathbb{I}(\cdot)\) is the indicator function, which returns 1 if the condition is satisfied and 0 otherwise, and \(\tau = 0.1\) is the chosen threshold.

As shown in the Supplementary Table~\ref{tab1}, TSENN consistently outperforms TSENN--S across all values of \(\ell_{\text{max}}\), despite TSENN--S using a significantly larger number of parameters. While the differences in \(\operatorname{MAE}\) are relatively small, the gap in \(\operatorname{MAE}_{\text{aniso}}^{\text{norm}}\) is substantial--often spanning orders of magnitude. Notably, TSENN--S, due to its scalar-only output, breaks the symmetry of the message-passing process and fails to capture the tensorial structure of the response. Even in the simplest case of cubic lattices, it cannot reproduce the clean isotropic behavior. Based on these results, we conclude that the optimal output representation is \(N_\omega \times 0e + N_\omega \times 2e\), for balancing spectral accuracy and anisotropy fidelity.
\color{black}

\begin{table}[ht]
\captionsetup{labelformat=empty} 
\caption{\textbf{Supplementary Table 1.} Comparison of TSENN models trained under the equivariant and scalar strategies.}
 
\centering
\label{tab1}
\begin{tabular*}{\textwidth}{@{\extracolsep\fill} c cccc cccc}
\toprule
& \multicolumn{4}{c}{TSENN (equivariant)} & \multicolumn{4}{c}{TSENN--Scalar } \\
\cmidrule(lr){2-5} \cmidrule(lr){6-9}
$\ell_{\text{max}}$ 
& \# params & $\text{MAE} \downarrow$ 
& $\text{MAE}_{\text{aniso}}^{\text{norm}} \downarrow$ & $\rho_{\text{aniso}}^{\text{norm}} \uparrow$
& \# params & $\text{MAE} \downarrow$ 
& $\text{MAE}_{\text{aniso}}^{\text{norm}} \downarrow$ & $\rho_{\text{aniso}}^{\text{norm}} \uparrow$ \\
\midrule
0 & 6.2M & 0.138 &  0.233 & 29.5\% & 8.2M  & 0.157 &  0.230 & 29.8\% \\
1 & 12.1M & 0.140 &  0.162 & 55.0\% & 29.3M & 0.158 &  0.231 & 29.7\% \\
2 & 25.5M & 0.130 &  0.143 & 71.5\% & 41.4M & 0.154 &  0.232 & 29.6\% \\
3 & 39.4M & 0.129 & 0.140 & 80.9\% & 55.3M & 0.160 &  0.229 & 29.9\% \\
4 & 55.4M & \textbf{0.125} & \textbf{0.139} & \textbf{81.6\%} & 71.3M & 0.166 &  0.241 & 21.6\% \\
\bottomrule
\end{tabular*}
\end{table}

\subsection{Loss Balancing for Multi-Objective Training}
Since the model simultaneously predicts both the \(\ell = 0\) (isotropic) and \(\ell = 2\) (anisotropic) components, the training objective naturally consists of two distinct loss terms. We observed that the magnitudes of these two losses differ significantly during training, with the loss associated with the \(\ell = 0\) component often being up to 10 times larger than that for the \(\ell = 2\) component. This imbalance can hinder the model's ability to effectively learn the anisotropic component. To address this problem, we apply an uncertainty-based loss balancing method during training, following the formulation proposed in~\cite{kendallMultiTaskLearningUsing2018}. We define the total loss function \(\mathcal{L}\) as being composed of two individual objectives \(\mathcal{L}_1\) and \(\mathcal{L}_2\), is defined as:
\begin{equation}
\mathcal{L} = \frac{1}{2 \sigma_1^2} \mathcal{L}_1(\ell=0) + \frac{1}{2 \sigma_2^2} \mathcal{L}_2(\ell=2) + \log \sigma_1 + \log \sigma_2,
\end{equation}
where \(\sigma_1\) and \(\sigma_2\) are learnable parameters that represent the relative uncertainty of each task. Minimizing \(\mathcal{L}\) with respect to both the model weights and \(\sigma_i\) allows the network to automatically learn an optimal weighting between the two objectives based on the data.

\begin{figure}[H]
    \centering
    \captionsetup{labelformat=empty}
    \includegraphics[width=0.7\linewidth]{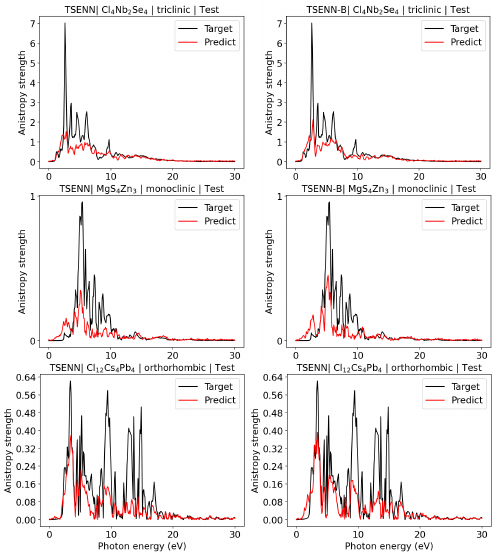}
    \caption{\textbf{Supplementary Figure 6.} Comparison of anisotropy strength predictions between TSENN and TSENN--B on three representative crystal systems from the test set. Each panel shows the Frobenius norm of the anisotropic component of the dielectric tensor as a function of photon energy, with target (black) and predicted (red) spectra. When the anisotropy strength is relatively large, both TSENN and TSENN--B reproduce the overall behavior well. The introduction of the balancer yields modest improvements, particularly near peak positions and in capturing finer tensorial details. These results suggest that uncertainty-based loss balancing can provide incremental gains in modeling anisotropic features within multi-objective training.}
    \label{SI_Plot_6}
\end{figure}

We refer to the model trained with the uncertainty-based loss balancing strategy as TSENN--Balancer (TSENN--B). To assess its performance on anisotropic responses, we examined three representative compounds with different levels of anisotropy: triclinic \ce{Cl4Nb2Se4} (mp-27361), monoclinic \ce{MgS4Zn3} (mp-1221971), and orthorhombic \ce{Cl12Cs4Pb4} (mp-675524). Supplementary Fig.~\ref{SI_Plot_6} shows the Frobenius norm of the anisotropic component of the dielectric tensor as a function of photon energy, comparing target spectra (black) with predictions (red). Both TSENN and TSENN--B reproduce the overall spectral behavior well. However, TSENN--B provides modest improvements, particularly near the peak positions. These results suggest that uncertainty-based loss balancing can incrementally enhance the model's ability to capture subtle anisotropic features in multi-objective training.

We further evaluate TSENN-B using the three metrics described above, across a range of $\ell_{\max }$ values; detailed results are provided in Supplementary Table~\ref{tab2}. As expected, TSENN-B performs comparably to TSENN on the overall spectral prediction task, with the best results achieved at $\ell_{\max }=4$. More importantly, TSENN-B consistently outperforms TSENN on the anisotropic component, as reflected in the lower associated values of $\text{MAE}_{\text{aniso}}^{\text{norm}}$ and higher $\rho_{\text{aniso}}^\text{norm}$. These results underscore the necessity of adaptive loss weighting when training models on multiobjective tasks with unbalanced loss magnitudes.
\begin{table}[ht]
\captionsetup{labelformat=empty} 
\caption{\textbf{Supplementary Table 2.} Comparison between TSENN models trained with the loss balancer (TSENN--Balancer) and without it (regular TSENN).}
 
\centering
\label{tab2}
\begin{tabular*}{\textwidth}{@{\extracolsep\fill} c cccc cccc}
\toprule
& \multicolumn{4}{c}{TSENN } & \multicolumn{4}{c}{TSENN--Balancer } \\
\cmidrule(lr){2-5} \cmidrule(lr){6-9}
$\ell_{\text{max}}$ 
& \# params & $\text{MAE} \downarrow$ 
& $\text{MAE}_{\text{aniso}}^{\text{norm}} \downarrow$ & $\rho_{\text{aniso}}^{\text{norm}} \uparrow$
& \# params & $\text{MAE} \downarrow$ 
& $\text{MAE}_{\text{aniso}}^{\text{norm}} \downarrow$ & $\rho_{\text{aniso}}^{\text{norm}} \uparrow$ \\
\midrule
0 & 6.2M & 0.138 &  0.233 & 29.5\% & 6.2M & 0.138 &  0.233 & 29.5\% \\
1 & 12.1M & 0.140 &  0.162 & 55.0\% & 12.1M & 0.135  & 0.162 & 57.1\% \\
2 & 25.5M & 0.130 &  0.143 & 71.5\% & 25.5M & 0.130 &  0.142 & 73.9\% \\
3 & 39.4M & 0.129 & 0.140 & 80.9\% & 39.4M & 0.129 &  0.141 & 81.5\% \\
4 & 55.4M & 0.125 & 0.139 & 81.6\% & 55.4M & \textbf{0.124} &  \textbf{0.138} & \textbf{82.8\%}\\
\bottomrule
\end{tabular*}
\end{table}

\subsection{Accuracy Contributions and Leakage Verifications}
 
To investigate the accuracy contributions from the two channels, and to check for possible leakage during training, we consider the decomposition of the prediction task into two essentially independent parts. Since the model output is bipartite, corresponding to the isotropic ($\ell=0$) and anisotropic ($\ell=2$) channels, one can, in principle, perform gradient descent on either channel alone and evaluate the magnitude of the other channel at convergence, or visualize its evolution during training. 

Supplementary Table~\ref{tab3} reports the mean absolute error calculated in the spherical--harmonics basis, which provides a natural separation into the invariant trace ($\ell=0$) and the traceless anisotropic part ($\ell=2$) of the dielectric tensor. Namely,  
\begin{equation}
    \text{MAE}^{\ell=0} = \frac{1}{N_\omega} 
    \sum_m \sum_\omega \big| \hat Y^0_m(\omega) - Y^0_m(\omega) \big|,
\end{equation}
\begin{equation}
    \text{MAE}^{\ell=2,m} = \frac{1}{N_\omega} 
     \sum_\omega \big| \hat Y^2_m(\omega) - Y^2_m(\omega) \big|,
\end{equation}
where $m$ runs over the five components of the $\ell=2$ subspace. The total MAE can then be written as  
\begin{equation}
    \text{MAE}^{\text{total}} 
    = \frac{1}{6}\Big(\text{MAE}^{\ell=0} + \sum_{m=-2}^2 \text{MAE}^{\ell=2,m}\Big),
\end{equation}
so that the channel-wise contributions are given by  
\begin{align}
    \text{Contrib}^{\ell=0} &= \tfrac{1}{6}\,\text{MAE}^{\ell=0}, \\
    \text{Contrib}^{\ell=2} &= \tfrac{1}{6}\sum_{m=-2}^{2} \text{MAE}^{\ell=2,m}.
\end{align}
By construction,
\begin{equation}
    \text{MAE}^{\text{total}} = \text{Contrib}^{\ell=0} + \text{Contrib}^{\ell=2},
\end{equation}
which allows us to quantify the relative error contributions of the isotropic and anisotropic channels while ensuring they sum exactly to the total. As shown in Supplementary Table~\ref{tab3}, when training on both channels together with cubic systems included, approximately 58.9\% of the error arises from the isotropic ($\ell=0$) channel and 41.1\% from the anisotropic ($\ell=2$) channel. This imbalance is expected, since cubic crystals contribute exclusively to the $\ell=0$ error. To address this bias, we repeated the analysis after removing cubic systems. In this case, the contributions become more balanced, with 52.2\% from $\ell=0$ and 47.8\% from $\ell=2$, confirming that both channels contribute comparably once symmetry constraints are taken into account.  

Furthermore, when training only on a single channel (either $\ell=0$ or $\ell=2$), the complementary channel vanishes and does not contribute to the overall error. This verifies that there is no leakage in our construction: the isotropic and anisotropic components represent distinct physical features and remain cleanly separated within the equivariant model.
\begin{table}[ht]
\captionsetup{labelformat=empty} 
\centering
\caption{ \textbf{Supplementary Table 3.} Ablation study on $\ell=0$ and $\ell=2$ channels. We compare training with both channels together vs. only one channel, and report mean MAE under the spherical--harmonics basis. Percentages denote relative contribution.}
\label{tab3}
\begin{tabular*}{\textwidth}{@{\extracolsep\fill} lcccc}
\toprule
Mode & $\text{MAE}^{\text{total}}$ & $\ell=0$ contrib. & $\ell=2$ contrib. & Cross-channel leakage \\
\midrule
Both (with cubic)     & 0.109 & 0.064 (58.9\%) & 0.045 (41.1\%) & -- \\
Both (without cubic)  & 0.132 & 0.069 (52.2\%) & 0.063 (47.8\%) & -- \\
$\ell=2$ only  & 0.068 & 0.000 (0.0\%)  & 0.068 (100\%) & no leakage to $\ell=0$ \\
$\ell=0$ only  & 0.075 & 0.075 (100\%) & 0.000 (0.0\%)  & no leakage to $\ell=2$ \\
\bottomrule
\end{tabular*}
\end{table}

Additionally, we analyzed the error distribution at the level of individual spherical--harmonic subchannels, reporting 
\(\text{MAE}^{\ell=0}\) and \(\text{MAE}^{\ell=2,m}\). 
To avoid statistical bias from symmetry-forbidden components, which would trivially yield zeros, we restrict the statistics to symmetry-allowed channels for each crystal system. 
For example, all six channels contribute in triclinic systems, while monoclinic systems allow \(\ell=0\), \(\ell=2,m=\pm2,\pm1,0\), with \(\ell=2,m=2\) corresponding to the \(xx-yy\) anisotropy, \(\ell=2,m=1\) to the off-diagonal \(xz+zx\) component, and \(\ell=2,m=0\) describing the traceless diagonal combination 
\(-\tfrac{1}{2}(xx + yy) + zz\).  
The normalized errors \(\text{NMAE}^{\ell,m}\) (\%) are defined relative to the characteristic magnitude of each block channel, providing a scale-aware comparison across irreps, see Supplementary Table~\ref{tab4}.  

We emphasize that this subchannel analysis offers a complementary view to Table~1 in the main text. While the trends are consistent, absolute values differ slightly due to the use of distinct characteristic magnitudes and coupling coefficients when performing the basis transformation. 
Interestingly, the largest errors arise in the \(\ell=2,m=0\) and \(\ell=2,m=2\) channels, consistent with the stronger diagonal anisotropies in the dielectric response. Nevertheless, all channels remain well balanced, with $\text{NMAE}^{\ell,m}$ below 10\%.  

\begin{table}[ht]
\captionsetup{labelformat=empty} 
\centering
\caption{\textbf{Supplementary Table 4. }Error analysis by spherical--harmonic subchannel. Mean values are reported with medians in parentheses.}
\label{tab4}
\begin{tabular*}{\textwidth}{@{\extracolsep\fill} lcc}
\toprule
Channel & $\text{MAE}^{\ell,m}$ & $\text{NMAE}^{\ell,m}$ (\%) \\
\midrule
\(\ell=0, m=0\) & 0.345 (0.254) & 2.9 (2.1) \\
\(\ell=2,m=-2\) & 0.129 (0.032) & 5.3 (1.3) \\
\(\ell=2,m=-1\) & 0.148 (0.130) & 6.1 (5.3) \\
\(\ell=2,m=0\)  & 0.185 (0.156) & 7.6 (6.4) \\
\(\ell=2,m=1\)  & 0.115 (0.101) & 4.7 (4.2) \\
\(\ell=2,m=2\)  & 0.199 (0.149) & 8.2 (6.1) \\
\bottomrule
\end{tabular*}
\end{table}
\color{black}

\newpage
\subsection{Basis Choices: Cartesian Versus Spherical Harmonics}
 
A central design choice in this work concerns how the dielectric tensor is represented. While both the Cartesian basis and the spherical--harmonic decomposition span the same number of independent variables--for example, two in uniaxial systems ($xx$, $zz$), four in monoclinic systems ($xx$, $xz$, $yy$, $zz$), and six in triclinic systems--the spherical basis offers distinct advantages that go beyond simple dimensionality. Its main strength lies in providing a symmetry-adapted, physics-inspired representation. In the Cartesian basis, rotational symmetry is not automatically enforced: as illustrated in Supplementary Fig.~\ref{fig:SI_Plot_7}, direct training in Cartesian components can lead to broken isotropy even in cubic crystals, where the exact relation $xx=yy=zz$ must hold. By contrast, spherical--harmonic decomposition enforces such relations by construction, ensuring that isotropic and anisotropic contributions remain cleanly separated and that cubic isotropy is preserved without additional constraints (see Supplementary Fig.~\ref{SI_Plot_3}).

An additional benefit of the spherical representation is robustness under distortions. Real materials rarely maintain perfect lattice symmetry, since structural relaxations, external strains, and thermal fluctuations can all induce distortions. In the Cartesian basis, tensor components mix under these perturbations in a complicated and non-transparent fashion, whereas in the spherical basis they transform naturally through the SO(3)-equivariant mixing of irreducible representations. This property makes the spherical representation both stable and interpretable, a feature we explicitly validated through shear-strain tests (see SI Sec.~\ref{SI:strain_verification}).

Finally, the spherical basis provides a more physically interpretable decomposition of the dielectric tensor. It separates the isotropic response ($\ell=0$) from the traceless tensorial anisotropy ($\ell=2$), yielding a framework that is directly aligned with the rotational symmetries of the problem and avoids leakage between channels. In this way, spherical--harmonic decomposition not only respects fundamental symmetry principles but also offers a transparent lens through which to analyze isotropic and anisotropic dielectric responses.

Taken together, these advantages highlight why the spherical--harmonic basis is particularly well-suited for learning tensorial dielectric spectra. Unlike Cartesian basis, which risk broken symmetries and obscured physical meaning, the spherical representation enforces correct invariances, adapts naturally to distortions, and provides an interpretable partitioning of isotropic and anisotropic contributions. Thus, while both bases span the same variable space, only the spherical--harmonics basis guarantees equivariance and physical consistency across all symmetry classes.

\color{black}

\begin{figure}[H]
    \centering
    \captionsetup{labelformat=empty}
    \includegraphics[width=0.9\linewidth]{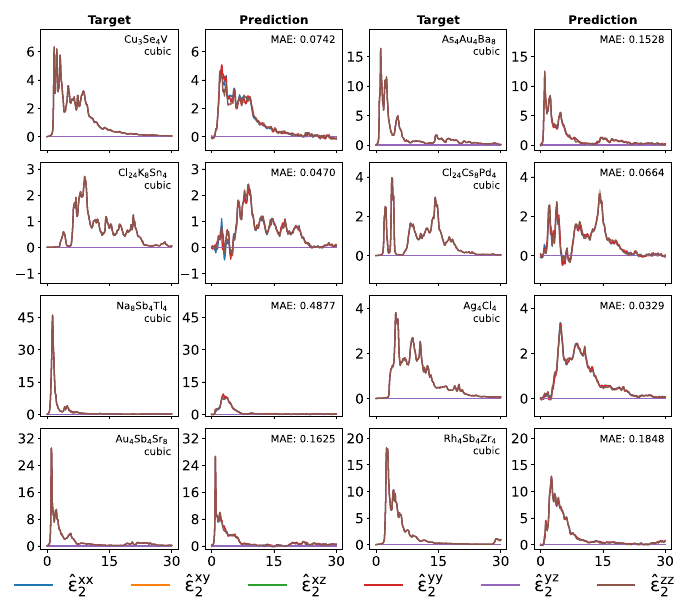}
    \caption{\textbf{Supplementary Figure 7.} 
    Comparison of target (left panels) and predicted (right panels) imaginary dielectric tensor spectra for selected cubic crystals. 
    In the target data, isotropy enforces $\hat{\varepsilon}_2^{xx} = \hat{\varepsilon}_2^{yy} = \hat{\varepsilon}_2^{zz}$, 
    resulting in a single line for the diagonal components (color-coded curves overlap perfectly). 
    By contrast, the predictions visibly separate into three distinct curves, indicating that isotropy is not preserved and 
    the equivariance constraint is broken. This discrepancy highlights the necessity of a spherical--harmonic representation 
    to enforce rotational symmetry and maintain consistency with cubic lattice constraints.}
    \label{fig:SI_Plot_7}
\end{figure}

\section{Comparison With Attention-Based Equivariant Models}
We extended our study by benchmarking against a state-of-the-art method for direct prediction of the full piezoelectric tensor: EATGNN~\cite{dongAccuratePiezoelectricTensor2025}, an attention-based equivariant graph neural network with an output irreps of $2 \times 1 o+1 \times 3 o+1 \times 4 o(\mathrm{e} 3 \mathrm{nn}$ notation). However, since our target differs significantly --being a rank-2 tensor with an additional photon-energy axis-we made minimal modifications to adapt the EATGNN architecture. We denote this adaptation of EATGNN as TSENN-Attention (TSENN--A). To ensure a direct comparison, we further adopted the same output representation, $N_\omega \times 0 e+N_\omega \times 2 e$, to obtain our model TSENN Balancer (TSENN--B), and applied uncertainty-based loss balancing to mitigate the scale differences between the scalar and tensor channels. Because our task does not involve irreducible representations with odd parity and the original architecture was designed for a different objective, we excluded all odd-parity irreps and performed a hyperparameter search with Optuna to determine the optimal even-parity representations. The best-performing attention configuration used $32 \times 0e + 16 \times 1e + 8 \times 2e + 4 \times 3e$ as the query/key irreps, while the output irreps were left unchanged. All other components of the architecture followed the original implementation.

Supplementary Table~\ref{tab5} compares the performance of TSENN--A and TSENN--B across different values of \(\ell_{\text{max}}\). In the \(\ell_{\text{max}} = 0\) case, TSENN--A functions as an invariant model, and its performance is comparable to that of TSENN--B. However, as \(\ell_{\text{max}}\) increases, TSENN--B consistently outperforms TSENN--A. This trend holds across all evaluation metrics, indicating that TSENN--B is more effective at capturing both the tensorial and spectral characteristics of the data.

\begin{figure}[H]
    \centering
    \captionsetup{labelformat=empty}
    \includegraphics[width=0.8\linewidth]{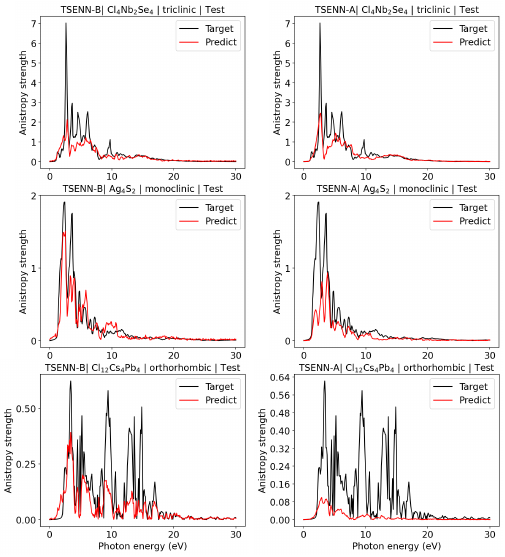}
    \caption{\textbf{Supplementary Figure 8.} Comparison between TSENN--B and TSENN--A is shown for three representative crystal systems from the test set, covering a range of anisotropic strengths: triclinic \ce{Cl4Nb2Se4} (strong anisotropy), monoclinic \ce{Ag4S2} (moderate anisotropy), and orthorhombic \ce{Cl12Cs4Pb4} (subtle anisotropy). While both models are equivariant and tensor-aware, TSENN--A performs well in the triclinic case but fails to capture the more subtle anisotropic behavior in the monoclinic and orthorhombic systems. In contrast, TSENN--B consistently recovers the anisotropic strength across all three cases. These results indicate that the attention-based construction of tensor interactions in TSENN--A may be less effective than the direct tensor product approach employed in TSENN--B, particularly for modeling subtle or highly anisotropic features.}
    \label{SI_Plot_8}
\end{figure}

Supplementary Fig.~\ref{SI_Plot_8} presents detailed predictions for three representative crystal systems. TSENN--A successfully captures the anisotropic response in the triclinic system but fails to do so in the monoclinic and orthorhombic cases. This suggests that the use of attention mechanisms does not significantly improve performance in capturing subtle anisotropic features.

A notable difference also arises in training efficiency. Both TSENN--B and TSENN--A were trained for 100 epochs on a single NVIDIA RTX 4090 GPU. However, because of the additional computational overhead of evaluating tensor products through attention mechanisms, TSENN--A required roughly four times longer to complete training. Despite having a comparable number of parameters, TSENN--A consistently underperforms relative to TSENN--B. This indicates that the attention-based construction of tensor interactions in TSENN--A is less effective than the direct tensor product approach used in TSENN--B, particularly for capturing subtle anisotropic behavior.

In summary, while attention mechanisms are commonly used to increase model expressiveness, our analysis indicates that they are not the primary factor driving performance in this task. Instead, the direct use of tensor products--combined with loss balancing between scalar and tensor channels--is more effective for learning both the underlying tensorial structure and the continuous spectral response.

\begin{table}[h]
\captionsetup{labelformat=empty} 
\caption{\textbf{Supplementary Table 5.} Comparison of two TSENN architectures: TSENN--Balancer (trained with the loss balancer but without attention) and TSENN--Attention (trained with both the loss balancer and the attention mechanism).}
\label{tab5}
 
\begin{tabular*}{\textwidth}{@{\extracolsep\fill} c cccc cccc}
\toprule
& \multicolumn{4}{c}{TSENN--Balancer} & \multicolumn{4}{c}{TSENN--Attention} \\
\cmidrule(lr){2-5} \cmidrule(lr){6-9}
$\ell_{\text{max}}$
& \#params & MAE $\downarrow$ & MAE$_{\text{aniso}}^{\text{norm}}$ $\downarrow$ & $\rho_{\text{aniso}}^{\text{norm}}$ $\uparrow$
& \#params & MAE $\downarrow$ & MAE$_{\text{aniso}}^{\text{norm}}$ $\downarrow$ & $\rho_{\text{aniso}}^{\text{norm}}$ $\uparrow$ \\
\midrule
0 & 6.2M  & 0.138  & 0.233 & 29.5\% & 27.2M  & 0.142  & 0.233 & 29.5\% \\
1 & 12.1M  & 0.135  & 0.162 & 57.1\% & 29.1M  & 0.139 &  0.172 & 47.8\% \\
2 & 25.5M & 0.130  & 0.142 & 73.9\% & 31.8M & 0.133 & 0.150 & 70.7\% \\
3 & 39.4M & 0.129  & 0.141 & 81.5\% & 33.0M & 0.131 & 0.150 & 71.2\%\\
4 & 55.4M & \textbf{0.124}  & \textbf{0.138} & \textbf{82.8\%} & 33.7M & 0.130 & 0.148 & 75.6\% \\
\bottomrule
\end{tabular*}
\end{table}

\newpage
\section{Benchmark Calculations for Dielectric and Optical Properties}
\begin{figure}[H]
    \centering
    \captionsetup{labelformat=empty}
    \includegraphics[width=0.9\linewidth]{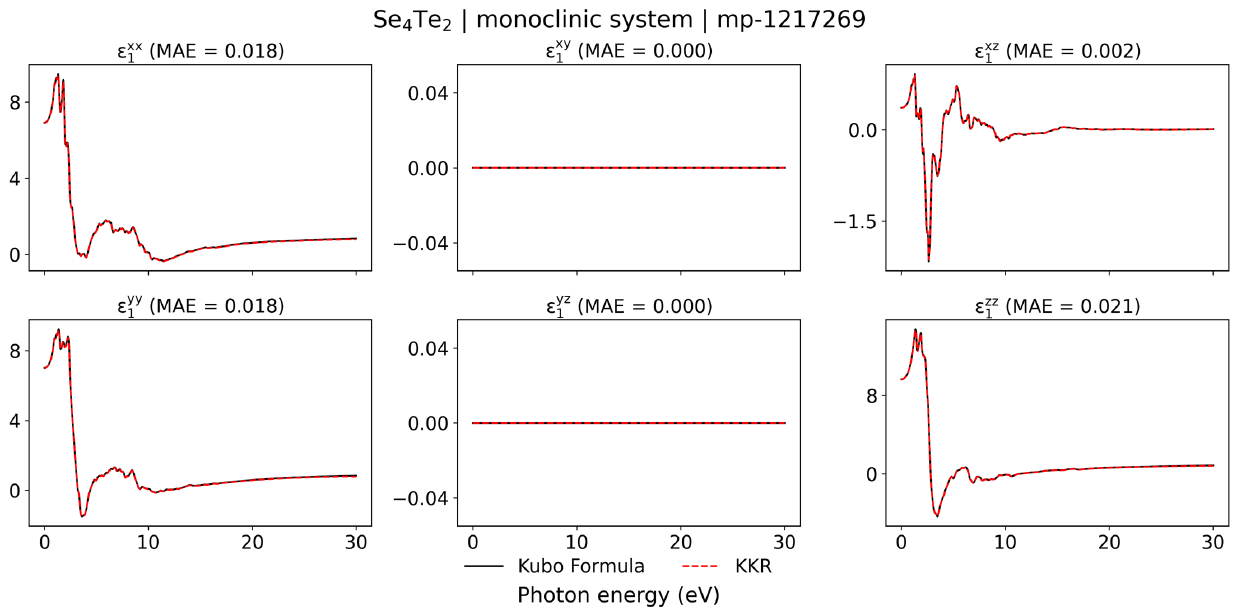}
    \caption{\textbf{Supplementary Figure 9.} Comparison between the real part of the dieletric tensor obtained via K-K relations from the computed imaginary spectra and the direct calculation using the Kubo formula. Results are shown for the monoclinic compound \ce{Se4Te2}, highlighting the four non-vanishing components: \(\varepsilon^{xx}\), \(\varepsilon^{yy}\), \(\varepsilon^{zz}\), and \(\varepsilon^{xz}\). Components such as \(\varepsilon^{xy}\) and \(\varepsilon^{yz}\) vanish due to symmetry constraints. The black line represents the direct Kubo calculation, while the red curve corresponds to the K-K relations reconstruction.}
    \label{fig:SI_Plot_9}
\end{figure}

In the main text, we primarily focused on the imaginary part of the dielectric tensor. However, the linear optical properties are intrinsically linked: the real part can be reconstructed numerically through the Kramers--Kronig (K-K) relation,
\begin{equation}
\varepsilon_1^{ij}(\omega) = 1 + \frac{2}{\pi} \,\fint_0^{\infty} 
\frac{ \varepsilon_2^{ij}(\omega')\,\omega'}{\omega'^2 - \omega^2}\,\mathrm{d}\omega',
\end{equation}
where $\fint$ denotes the Cauchy principal value. Once the real part is reconstructed, one can further derive related quantities such as the directional absorption coefficient, optical conductivity, and the complex refractive index. To assess the accuracy of this reconstruction, we consider the representative example of the monoclinic compound \ce{Se4Te2}. Starting from the computed imaginary part of the dieletric spectrum, we apply the K-K relation to obtain the real part and compare it against the corresponding direct computation based on the Kubo formula. This serves as a self-consistency check of our numerical integral and data-driven framework. Supplementary Fig.~\ref{fig:SI_Plot_9} shows that the reconstructed and directly computed real parts exhibit excellent agreement across all symmetry-allowed components with only minor discrepancies and lead to a small mean absolute error. The residual differences arise primarily from numerical approximations involved in the discretized integral: note that this integration becomes numerically feasible in our case because the imaginary part of the dieletric spectrum decays smoothly and approaches zero as \(\hbar\omega \to 30\)~eV, ensuring convergence of the finite upper limit approximation. Overall, this comparison confirms the internal consistency and robustness of our implementation.

In the main text, we demonstrate that the predicted spectra can be used to reconstruct the real part of the dieletric tensor to obtain the full complex-valued, frequency-dependent tensorial response. Note that the example provided, \ce{Se4Te2}, belongs to the test set and it was not included in the training set. This points to the wider applicability of our model and its predictive capabilities beyond the imaginary part of the dieletric tensor alone. To further validate this capability, we consider another monoclinic compound from the test set, \ce{K4O8Sb4}, and use the model to predict its imaginary spectrum. The reconstructed real part is then compared with the directly computed results obtained from the Kubo formula. As shown in Supplementary Fig.~\ref{fig:SI_Plot_10}, the agreement between the two is again found to be excellent, with the mean absolute errors of individual tensor components reported to quantify the accuracy. This level of precision should enable reliable reconstruction of physical observables derived from spectral properties with high fidelity.

\begin{figure}[H]
    \centering
    \captionsetup{labelformat=empty}
    \includegraphics[width=0.8\linewidth]{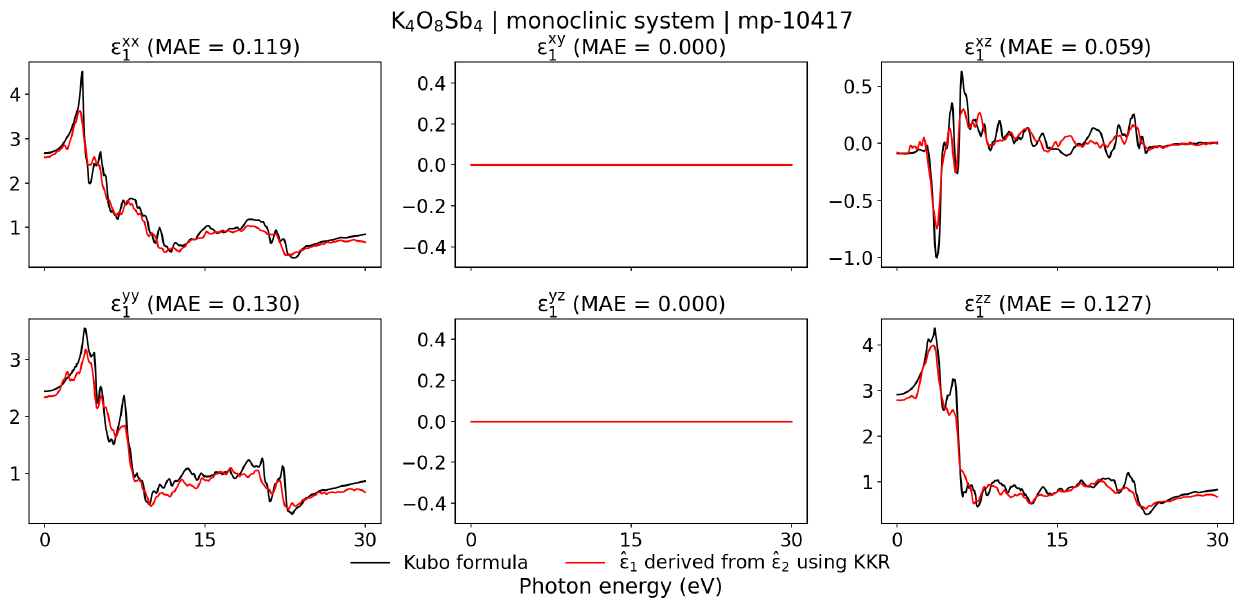}
    \caption{\textbf{Supplementary Figure 10.} Comparison between the real part of the dieletric tensor obtained via K-K relations from the model-predicted imaginary spectra and the direct calculation using the Kubo formula. Excellent agreement is observed across all non-vanishing components. Components such as \(\varepsilon^{xy}\) and \(\varepsilon^{yz}\) remain zero due to symmetry and thus do not contribute to the overall MAE.}
    \label{fig:SI_Plot_10}
\end{figure}
 
Accurately predicting the bandgap has long been a central challenge in materials research. It is particularly valuable to assess whether the bandgap can be inferred directly from predicted dielectric tensor. From the full complex dielectric tensor, we extract the absorption coefficients and, using the Tauc method~\cite{viezbickeEvaluationTaucMethod2015}, determine the corresponding optical bandgaps. By definition, the optical bandgap is the lowest photon energy at which an allowed electronic transition occurs. If direct transitions are forbidden at the electronic band edge, the optical bandgap will necessarily exceed the fundamental electronic bandgap. This explains why optical bandgaps are generally greater than or equal to the electronic bandgaps, making the comparison between the two physically consistent.  

The left panel of Supplementary Fig.~\ref{fig:SI_Plot_11} compares DFT electronic bandgaps with optical bandgaps obtained from Tauc fits to the calculated absorption functions, showing the expected systematic shift. The right panel compares optical bandgaps extracted from ML-predicted spectra with reference Tauc-derived values, achieving an $R^2$ of 0.897 on the test set. These results demonstrate that our model not only reproduces the imaginary dielectric spectra with high fidelity but also encodes sufficient information to recover the full dielectric function, and consequently absorption spectra and bandgaps, with high accuracy.  

\begin{figure}[H]
    \centering
    \captionsetup{labelformat=empty}
    \includegraphics[width=0.9\linewidth]{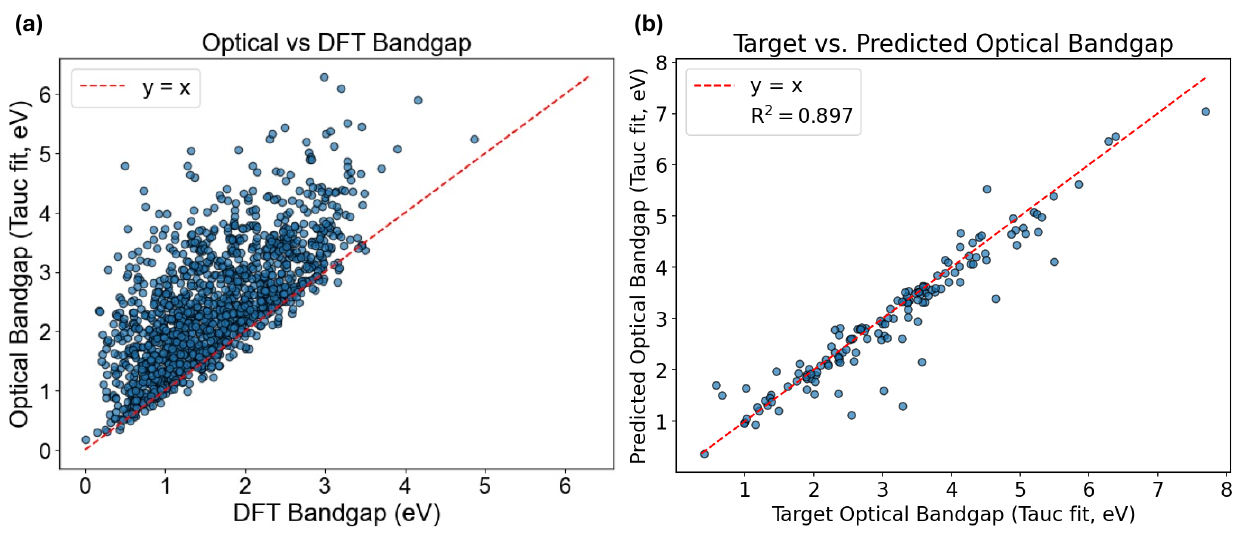}
    \caption{\textbf{Supplementary Figure 11.}
    \textbf{(a)} Comparison of DFT-calculated electronic bandgaps with optical bandgaps obtained from Tauc fits. 
    \textbf{(b)} Comparison of target Tauc-derived optical bandgaps with values extracted from ML-predicted absorption spectra, showing good agreement ($R^2 = 0.897$).}
    \label{fig:SI_Plot_11}
\end{figure}

An important aspect of the dielectric response is the role of the plasma frequency, $\omega_p$, which marks the transition between collective electronic oscillations and the asymptotic high-frequency regime. By reconstructing the full dielectric function, we extract $\omega_p$ from the energy-loss spectra~\cite{dealmeidaElectronicOpticalProperties2006} and use it as a natural divider of the spectral domain. Partitioning the benchmarking at $\omega_p$ provides a clearer view of how model accuracy depends on spectral regime. 

Below $\omega_p$, where material-specific absorption features with strong oscillator strengths dominate, the model exhibits substantially larger errors (diagonal MAE $\sim 0.489$--$0.532$; off diagonal MAE $\sim 0.146$--$0.183$; NMAE $\sim 6.5$\%--$13.3$\%). In contrast, above $\omega_p$ the spectra are smoother and less discriminative, and the errors drop by nearly an order of magnitude (diagonal MAE $\sim 0.044$--$0.05$; off diagonal MAE $\sim 0.012$--$0.015$; NMAE $\sim 0.6$\%--$1.1$\%), see Supplementary Fig.~\ref{fig:SI_Plot_12}. Compared to the aggregate errors reported in Table~1 of the main text, this frequency-resolved benchmarking reveals nearly a twofold increase in the measured errors within the absorption regime, underscoring that the low-frequency region provides the most meaningful and physically relevant assessment of model fidelity. 

It is worth noting that, although one could in principle introduce an energy-resolved weighting during training to emphasize $\omega < \omega_p$, such an approach requires prior knowledge of the plasma frequency for each material and would introduce material-dependent biases at inference time. Because $\omega_p$ varies widely across compounds, this would compromise the model's generalizability. For this reason, we adopted an energy-independent training loss and instead evaluated accuracy using frequency-resolved metrics, which we view as a more robust and transferable approach.

\begin{figure}[H]
    \centering
    \captionsetup{labelformat=empty}
    \includegraphics[width=0.7\linewidth]{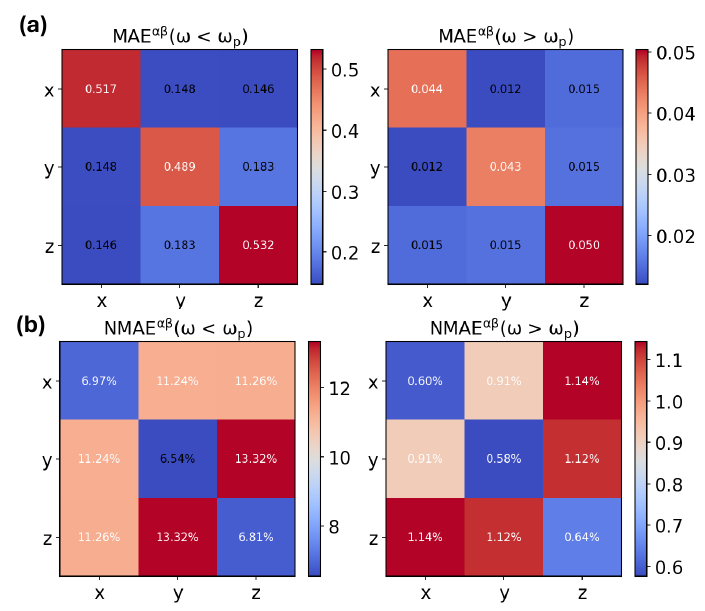}
    \caption{\textbf{Supplementary Figure 12.}
    Per-component $\text{MAE}^{\alpha\beta}$ and $\text{NMAE}^{\alpha\beta}$ of the imaginary dielectric tensor, evaluated separately below and above the plasma frequency $\omega_p$. 
    Once the spectra are partitioned at $\omega_p$, the errors become nearly twice as large compared to the aggregate values reported in Table~1 of the main text. 
    The low-frequency regime ($\omega < \omega_p$) dominates the error, with significant contributions from both diagonal and off-diagonal components, whereas the high-frequency regime ($\omega > \omega_p$) exhibits substantially smaller errors, with $\text{NMAE}^{\alpha\beta}$ reduced to only a few percent per component. 
    This trend is consistent with the physical picture: below $\omega_p$, the dielectric response captures rich, material-specific absorption features with pronounced oscillator strengths, making this regime intrinsically more sensitive to predictive errors, while the smoother high-frequency response is easier to approximate.}
    \label{fig:SI_Plot_12}
\end{figure}

\newpage
\section{Strain Verification} \label{SI:strain_verification}
To examine the preservation of equivariance, we consider two representative perturbations that probe the model's response to broken symmetry. First, we apply uniaxial strain along the $c$-axis to a cubic crystal in the test set, \ce{AgCl} (mp-22922). We compare the model-predicted dielectric tensor with results from explicit \emph{ab initio} calculations on the strained structures. Both show an identical linear dependence on the applied uniaxial strain (0--5\%), confirming that the model faithfully reproduces the expected anisotropic response.  

\begin{figure}[H]
    \centering
    \captionsetup{labelformat=empty}
    \includegraphics[width=0.9\linewidth]{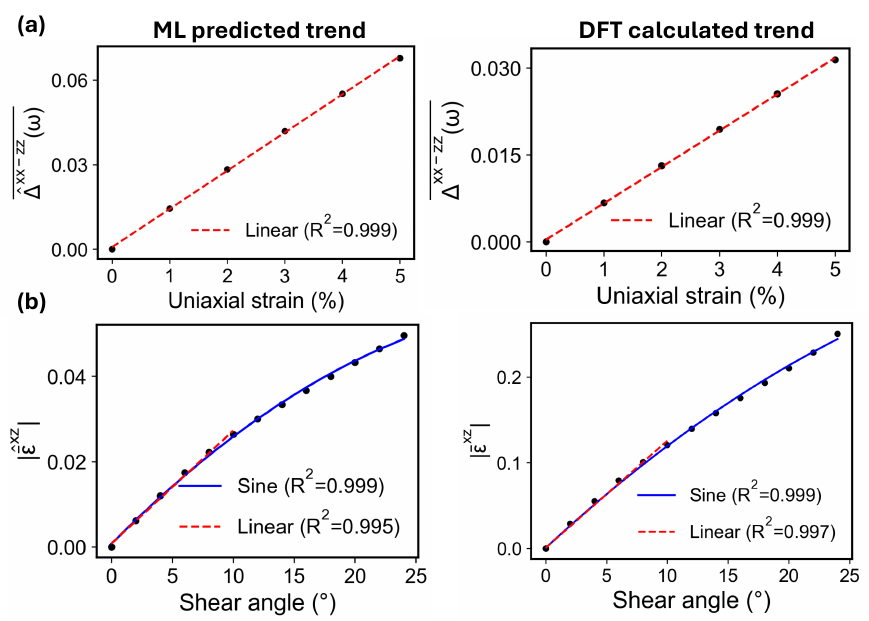}
    \caption{\textbf{Supplementary Figure 13.}
    Benchmarking of model predictions against \emph{ab initio} dielectric tensors under symmetry-breaking perturbations. 
    (\textbf{a}) Uniaxial strain (0--5\%) applied along the $c$-axis of \ce{AgCl}, showing a linear dependence of anisotropy mean along the frequency, namely, $\overline{\Delta^{xx-zz}(\omega)}$ captured equally by the model and \emph{ab initio} calculations. 
    (\textbf{b}) Shear distortions: for small shear angles ($<10^\circ$), the response is nearly linear, while for extended angles ($10^\circ \leq \theta < 25^\circ$) the dependence follows a sinusoidal trend. 
    In both cases, the model reproduces the expected functional form with high $R^2$ values, confirming its ability to capture anisotropic responses under strain and shear perturbations.}
    \label{fig:SI_Plot_13}
\end{figure}

Second, we analyze shear distortions in an orthorhombic crystal, \ce{Ag3AsS4} (mp-9538), where the response depends on the shear angle $\theta$. For small shear angles ($<10^\circ$), both model and \emph{ab initio} results follow a linear dependence, consistent with the Taylor expansion $\sin \theta \approx \theta$. At larger shear angles ($10^\circ \leq \theta < 25^\circ$), the response exhibits a sinusoidal trend, as physically expected. However, such large distortions are structurally unstable and energetically unfavorable in realistic crystals. To clarify this distinction, we present Supplementary Figure~\ref{fig:SI_Plot_13}, which compares (i) linear fits in the small-angle regime, corresponding to realistic structural deformations, and (ii) sine fits across the full angular range, capturing the asymptotic response under artificially large distortions. In both regimes, the model closely follows the trends obtained from \emph{ab initio} calculations, reproducing the linear-to-sinusoidal crossover with high fidelity. This agreement demonstrates the robustness of the model in capturing anisotropic responses under symmetry-breaking perturbations.

\color{black}

\section{Details of Cartesian Tensor Decomposition}
To analyze and manipulate tensorial responses in a symmetry-adapted manner, it is often advantageous to express Cartesian tensors in terms of spherical--harmonics. This transformation not only reveals the underlying angular momentum structure but also enables a natural decomposition into irreducible tensor components, making it readily compatible with the \texttt{e3nn} framework. Below, we present the formalism for decomposing a rank-2 Cartesian tensor into its spherical--harmonic components using a vector basis transformation and Wigner \(3j\)-symbols.

The Cartesian components of a unit vector can be expressed on the unit sphere as
\begin{equation}
\begin{aligned}
v_x &= \sin \theta \cos \phi, \\
v_y &= \sin \theta \sin \phi, \\
v_z &= \cos \theta.
\end{aligned}
\end{equation}
To connect Cartesian coordinates \(v_i\) with \(i\in \{x,y,z\}\) and the related spherical coordinates, we define the covariant components \(v_\alpha\) with \(\alpha \in \{-, 0, +\}\) via a linear transformation:
\begin{equation}
v_\alpha = C_{i\alpha} v^i,
\end{equation}
where the transformation matrix \(C_{i \alpha}\) is given by
\begin{equation}
C_{i \alpha} = \begin{pmatrix}
\frac{1}{\sqrt{2}} & 0 & -\frac{1}{\sqrt{2}} \\
-\frac{i}{\sqrt{2}} & 0 & -\frac{i}{\sqrt{2}} \\
0 & 1 & 0
\end{pmatrix}.
\end{equation}
From this, the covariant components can be explicitly expressed as
\begin{equation}
\begin{aligned}
v_{-} &= \frac{1}{\sqrt{2}}(v_x - i v_y), \\
v_{0} &= v_z, \\
v_{+} &= -\frac{1}{\sqrt{2}}(v_x + i v_y).
\end{aligned}
\end{equation}
The components \(v_\alpha\) can now be directly related to the spherical--harmonics \(Y_1^\alpha\), as both form a basis for angular momentum eigenstates in the \((-, 0, +)\) representation. More concisely, we can write:
\begin{equation}
v^i = C_{i\alpha}^* v_\alpha = C_{i\alpha}^* Y_1^\alpha.
\end{equation}
A crucial step in this transformation is the decomposition of tensor products of Cartesian coordinates into spherical--harmonics. Specifically, the rank-2 tensor \(v_i v_j\) can be expressed in the spherical basis as:
\begin{equation}
v^i v^j = C^*_{i\alpha} C^*_{j\beta} v_\alpha v_\beta = C^*_{i\alpha} C^*_{j\beta} Y_1^\alpha Y_1^\beta.
\end{equation}
We next evaluate the product \(Y_1^\alpha Y_1^\beta\) using Wigner \(3j\)-symbols:
\begin{equation}
\begin{aligned}
v_\alpha v_\beta = Y_1^\alpha Y_1^\beta 
=& \left(\begin{array}{ccc}
0 & 1 & 1 \\
0 & 0 & 0
\end{array}\right)
\left(\begin{array}{ccc}
0 & 1 & 1 \\
-(\alpha+\beta) & \alpha & \beta
\end{array}\right) Y_0^0 \\
& + 5(-1)^{\alpha+\beta}
\left(\begin{array}{ccc}
2 & 1 & 1 \\
0 & 0 & 0
\end{array}\right)
\left(\begin{array}{ccc}
2 & 1 & 1 \\
-(\alpha+\beta) & \alpha & \beta
\end{array}\right) Y_2^{\alpha+\beta}.
\end{aligned}
\end{equation}
The selection rules for Wigner \(3j\)-symbols \(\left(\begin{array}{ccc}
l_1 & l_2 & l_3 \\
m_1 & m_2 & m_3
\end{array}\right)\) requires that the total angular momentum \(\ell\) satisfies \(|\ell_1 - \ell_2| \leq \ell \leq \ell_1 + \ell_2\) and \(\ell_1 + \ell_2 + \ell\) must be even. Since \(\ell_1 = \ell_2 = 1\), only \(\ell = 0\) and \(\ell = 2\) are allowed. This leads to the concise decomposition:
\begin{equation}
v_\alpha v_\beta = Y_1^{\alpha} Y_1^\beta = Y_0^{\alpha+\beta} \oplus Y_2^{\alpha+\beta},
\end{equation}
which explains why only \(\ell = 0\) and \(\ell = 2\) components appear in the spherical--harmonic decomposition of Cartesian rank-2 tensors. 

We now examine the dielectric tensor, a rank-2 tensor denoted by \(\varepsilon^{ij}\). This tensor may be expressed in terms of the Cartesian tensor basis \(v^i v^j\) with the corresponding coefficients \(T^{ij}\). To transition to the spherical--harmonic basis, we apply the basis transformation via the coefficients \(C_{i\alpha}\) which yields:
\begin{equation}
\varepsilon^{ij} = T^{ij} v^i v^j = T^{ij} C_{i \alpha}^* C_{j \beta}^* Y_1^\alpha Y_1^\beta = T^{\alpha \beta} Y_1^\alpha Y_1^\beta.
\end{equation}
Once the transformation between the Cartesian and spherical--harmonic basis is established, the physical tensor can be projected onto the irreducible components of the spherical basis. This decomposition naturally separates the tensor into its scalar and quadrupolar parts, corresponding to the angular momentum channels \(\ell = 0\) and \(\ell = 2\), respectively:
\begin{equation}
\varepsilon^{ij} = T^{\alpha \beta} Y_1^\alpha Y_1^\beta = \varepsilon^{(0)} \oplus \varepsilon^{(2)}.
\end{equation}

\section{Data Downsampling and Balanced Splitting for Various Crystal Systems}\label{sec:data_splitting}
\begin{figure}[H]
    \centering
    \captionsetup{labelformat=empty}
    \includegraphics[width=0.8\linewidth]{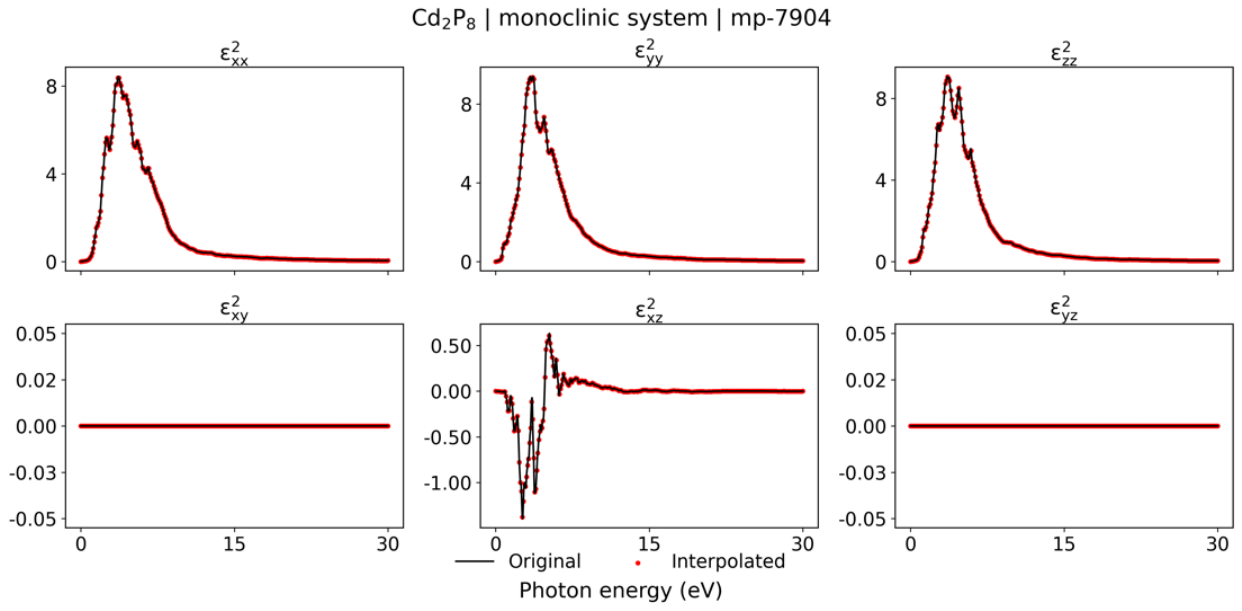}
    \caption{\textbf{Supplementary Figure 14.} Comparison between the original and interpolated imaginary parts of the frequency-dependent dielectric tensor components for \ce{Cd2P8}, a monoclinic crystal. Interpolated data (red dots) closely follow the original spectra (black dashed lines) and accurately capture peak features across both the diagonal and off-diagonal components. This demonstrates that the interpolation scheme effectively preserves the essential physical characteristics of the optical response across the full tensor.}
    \captionsetup{labelformat=default}
    \label{fig:SI_Plot_14}
\end{figure}

The original optical spectra were calculated at an interval of 0.01\,eV, resulting in 3001 data points spanning over 0--30\,eV. While this high-resolution sampling effectively captures fine spectral features, it also introduces significant computational overhead for both model training and evaluation using the K-K relations. To mitigate this, we applied a downsampling strategy inspired by signal processing techniques, reducing the number of data points to 300 while preserving the essential spectral information. As an illustrative example, we consider the monoclinic compound \ce{Cd2P8}, whose symmetry allows for non-vanishing components in the \(xx\), \(yy\), \(zz\), and \(xz\) elements of the dielectric tensor. Supplementary Fig.~\ref{fig:SI_Plot_14} shows that the downsampled spectra correctly preserve the key peak positions across all components, demonstrating that the essential optical features are preserved and remain suitable for training the model and reconstructing the spectra with high fidelity.
We observed that when certain crystal symmetries are missing from the training set, while this does not necessarily lead to unphysical predictions, it might limit the model's ability to generalize symmetry-consistent behavior across diverse crystal classes. Since our primary interest lies in capturing symmetry-dependent trends across different crystal systems, a balanced sampling strategy based on crystal symmetry is more appropriate than a split based solely on chemical species. To this end, we adopt a stratified data splitting approach in which the training, validation, and test sets are constructed to ensure that all crystal systems are adequately represented, see Supplementary Fig.~\ref{fig:SI_Plot_15}. Specifically, we split the dataset using an 80:10:10 ratio, assigning 80\% of the data to training, and 10\% each to validation and testing. This guarantee that every symmetry class appears in all subsets, allowing the model to encounter diverse crystal systems during training and evaluation. Moreover, since the validation set contains examples from all crystal systems, model selection based on validation performance inherently accounts for symmetry diversity. The best-performing model on this set is thus expected to generalize well, having been evaluated across a representative range of symmetry cases. This setup ensures a more robust and symmetry-aware training-validation workflow.

\begin{figure}[H]
    \centering
    \captionsetup{labelformat=empty}
    \includegraphics[width=0.8\linewidth]{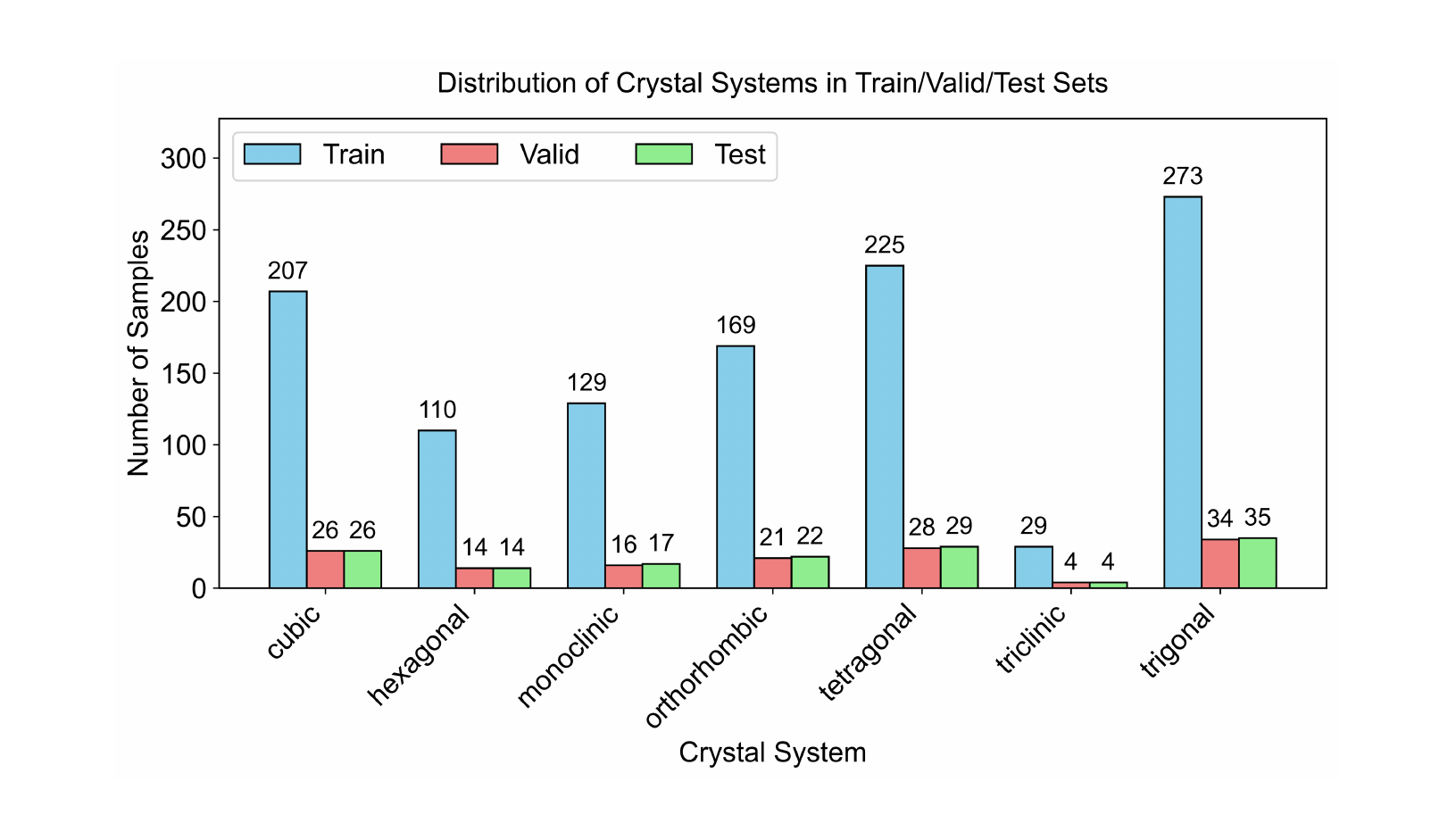}
    \caption{\textbf{Supplementary Figure 15.} Distribution of crystal systems across the training, validation, and test sets. Each bar indicates the number of samples from a given crystal system in the training (blue), validation (red), and testing (green) sets. By ensuring that all symmetry classes are represented across all subsets, this strategy facilitates more robust learning of symmetry-aware features and improves generalization to materials not included in the training set.}
    \label{fig:SI_Plot_15}
\end{figure}

\section{Focused Benchmarking on Low-Symmetry Anisotropic Materials}
To more rigorously evaluate the model's ability to capture anisotropic tensorial responses, we construct an auxiliary benchmark focused on low-symmetry materials. Our primary dataset contains few triclinic and monoclinic structures, so we broadened the selection criteria to increase their representation of these structure. Specifically, we restricted the crystal systems to triclinic and monoclinic structures and relaxed the filters to a band gap of 0.3--5 eV, an energy-above-hull range of 0--0.1 eV, compositions with 1--3 elements, and number of atomic sites from 1 to 40. The larger energy-above-hull window and allowance for bigger unit cells are of key importance here because many low-symmetry materials are metastable or possess larger primitive cells, and are therefore excluded in our main dataset. These relaxed constraints improve the coverage of low-symmetry systems but alter the structural and chemical distributions. For this reason, the auxiliary dataset should not be mixed with the original benchmark, as the auxiliary dataset contains more metastable structures and differs systematically in symmetry, complexity, and size.

Based on the OpenMX computational parameters used in the Methods section of the main text, we obtained self-consistent convergence for 1,658 structures, consisting of 1,177 monoclinic systems and 481 triclinic systems. We then adopted the same data splitting strategy discussed in SI Sec. \ref{sec:data_splitting} to train a model dedicated to materials with finite off-diagonal dielectric components. In this way, we obtained a more precise assessment of the efficacy of our model for predicting anisotropic components of tensorial optical spectra.

Supplementary Table \ref{tab6} summarized the performance of our model on the auxiliary dataset. Relative to the results in Table~1 of the main text, the auxiliary dataset yields similar overall performance but exhibits substantial improvements in the off-diagonal metrics, including reduced \(\text{MAE}^{\alpha\beta}\) and \(\text{NMAE}^{\alpha\beta} (\%)\). Per-structure prediction accuracy stratified by the \(\text{MAE}\) is visualized in Supplementary Fig.~\ref{fig:SI_Plot_16}, and by the \(\text{MAE}_{\text{aniso}}^{\text{norm}}\) distributions in Supplementary Fig.~\ref{fig:SI_Plot_17}. The off-diagonal components are predicted significantly more accurately than the one in the main dataset. This improvement is expected, as the auxiliary benchmark effectively acts as a data augmentation step trained exclusively on materials that exhibit non-zero anisotropy. These results demonstrate that, when provided with an appropriately anisotropy-rich dataset, the model can reliably learn the underlying symmetry relationships encoded in the crystal structures and map them to the corresponding tensorial spectra.

\begin{table}[ht]
\captionsetup{labelformat=empty} 
\caption{\textbf{Supplementary Table 6.} Summary of full-tensor and per-component error metrics for dielectric tensor predictions on the auxiliary dataset (mean values with median in parentheses).}
\label{tab6}
\begin{tabular*}{\textwidth}{@{\extracolsep\fill}cc|cccccc}
\toprule
\multicolumn{2}{c|}{\textbf{Full-tensor metrics}} & \multicolumn{5}{c}{\textbf{Per-component metrics }} \\
\midrule
Metric & Value & $\alpha\beta$ & $\mathrm{MAE}^{\alpha \beta}$ & $\mathrm{NMAE}^{\alpha \beta}$ (\%)  & $\mathrm{MAE}^{\prime, \alpha \beta}$ & KL divergence & $\text{MAE}_{\text{aniso}}^{\alpha\beta}$ \\
\midrule
MAE                         & 0.102 (0.091) & $xx$ & 0.172 (0.150) & 3.3 (3.4)& 0.071 (0.065) & 0.091 (0.044) & 0.094 (0.075) \\
$\text{MAE}_{\text{aniso}}$ & 0.084 (0.072) & $yy$ & 0.171 (0.148) & 3.4 (3.5)& 0.070 (0.063) & 0.093 (0.043) & 0.092 (0.081) \\
$\text{MAE}_{\text{aniso}}^{\text{norm}}$ & 0.146 (0.121) & $zz$ & 0.175 (0.150)& 3.4 (3.5) & 0.071 (0.066) & 0.110 (0.040) & 0.091 (0.099) \\
\cmidrule(lr){3-8}
 & & $xy^\dagger$ & 0.046 (0.041) & 5.8 (7.7) & 0.033 (0.031) & -- & 0.046 (0.041) \\
 & & $xz^\dagger$ & 0.065 (0.054) & 5.9 (7.3) & 0.025 (0.024) & -- & 0.065 (0.054) \\
 & & $yz^\dagger$ & 0.056 (0.055) & 5.7 (6.9) & 0.030 (0.026) & -- & 0.056 (0.055) \\
\botrule
\end{tabular*}
\begin{flushleft}
\footnotesize{$^\dagger$ Off-diagonal values are reported only for symmetry-allowed systems. 
Since the isotropic trace contributes only to diagonal elements, 
$\text{MAE}^{\alpha\beta} = \text{MAE}_{\text{aniso}}^{\alpha\beta}$ for off-diagonal terms. 
The KL divergence is not reported for off-diagonal components because these spectra are not positive definite, 
and the normalization required for KL is ill-defined.}
\end{flushleft}
\end{table}

\begin{figure}[H]
    \centering
    \captionsetup{labelformat=empty}
    \includegraphics[width=0.9\linewidth]{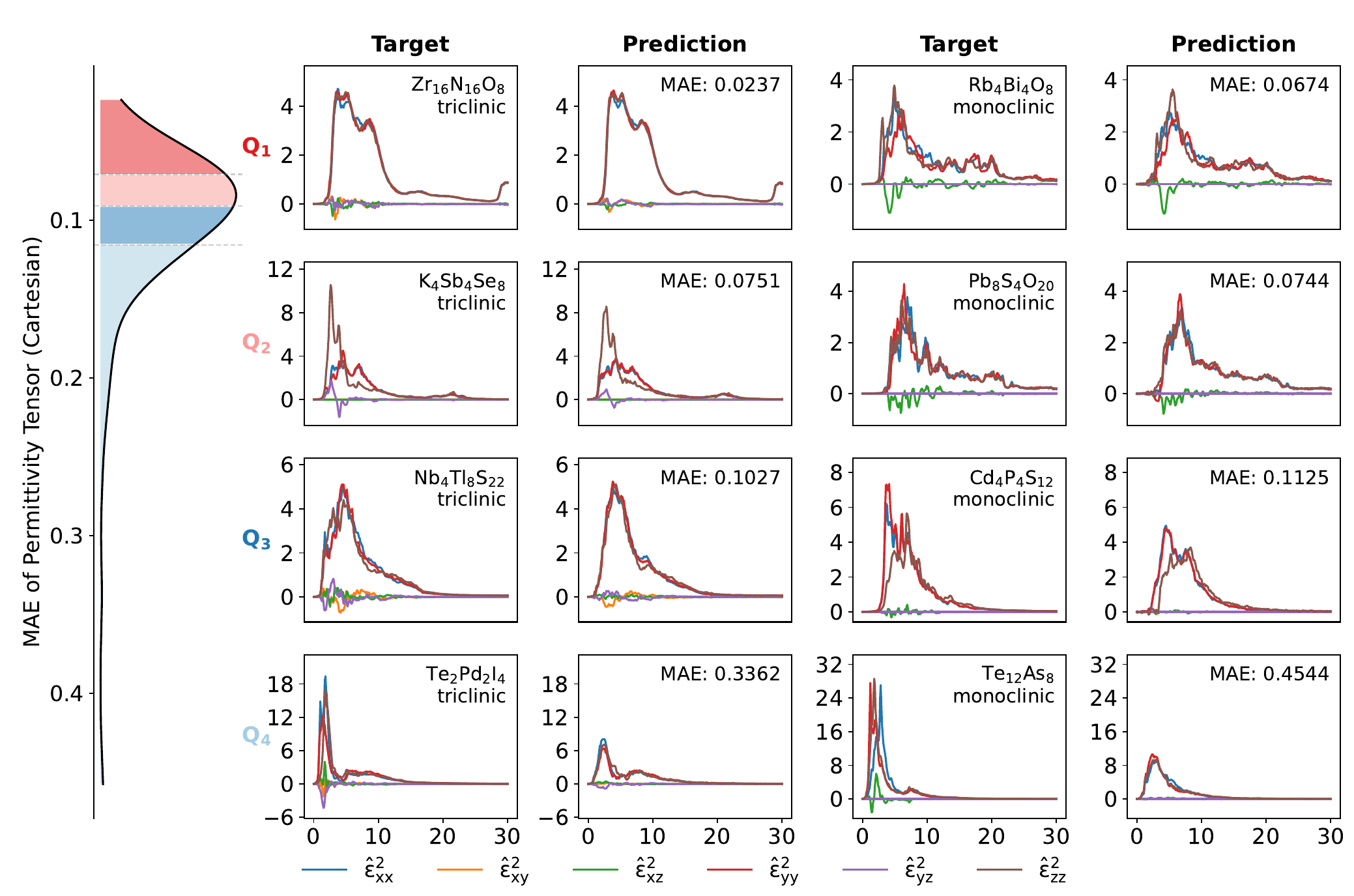}
    \caption{\textbf{Supplementary Figure 16.} Distribution of full-tensor MAE values across the test set of the auxiliary dataset (left) and representative spectra obtained by stratified sampling across quartiles (right). From the cumulative kernel-density-estimator (KDE) plot, we randomly selected two systems in each quartile ($Q_1$--$Q_4$), resulting in 8 representative examples that span the full error distribution.}
    \label{fig:SI_Plot_16}
\end{figure}

\begin{figure}[H]
    \centering
    \captionsetup{labelformat=empty}
    \includegraphics[width=0.9\linewidth]{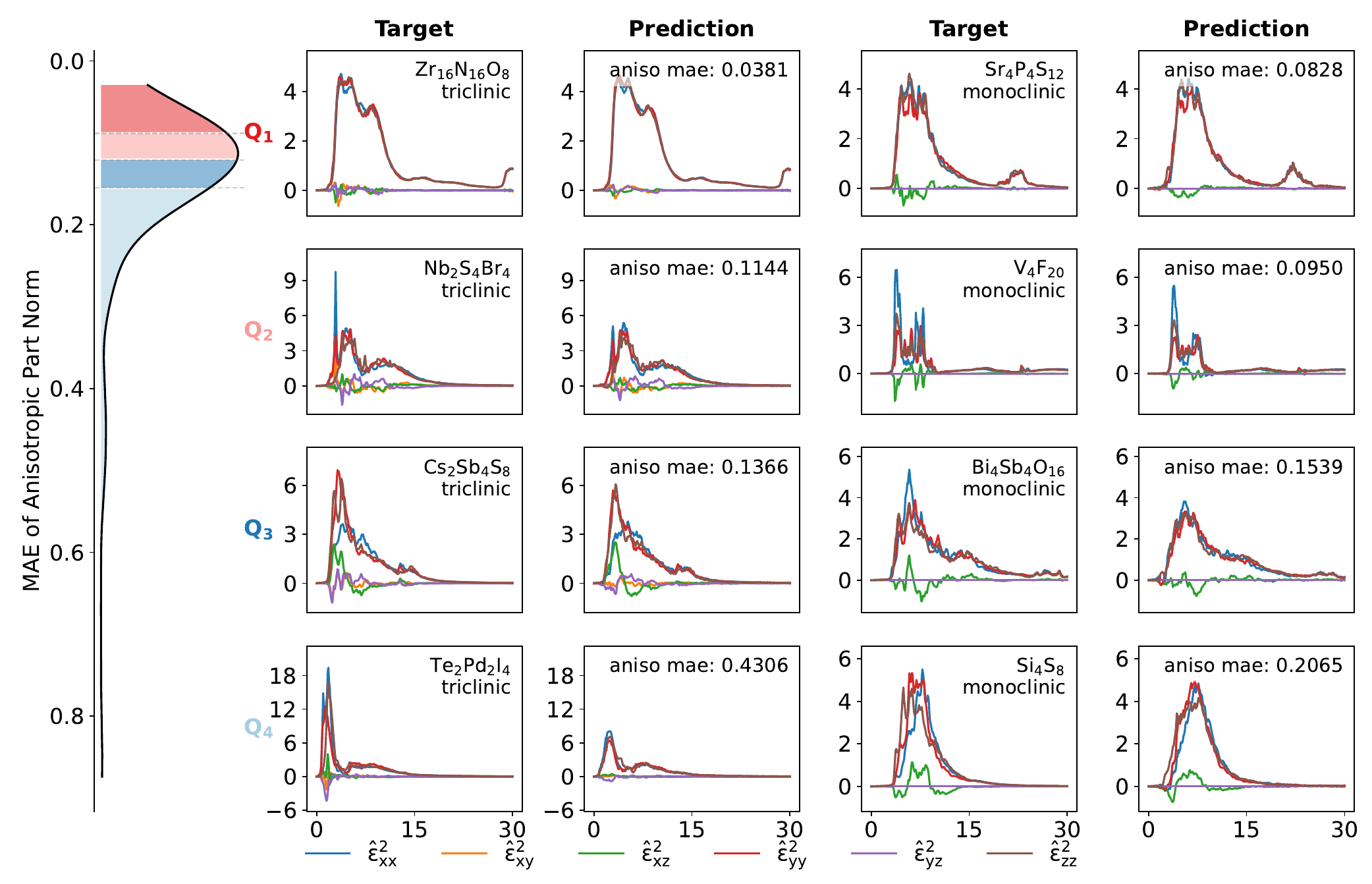}
    \caption{\textbf{Supplementary Figure 17.} Distribution of $\text{MAE}_{\text{aniso}}^{\text{norm}}$ values across the test set of the auxiliary dataset (left) and representative spectra obtained by stratified sampling across quartiles (right). From the cumulative kernel-density-estimator (KDE) plot, we randomly selected two systems in each quartile ($Q_1$--$Q_4$), resulting in 8 representative examples that span the full error distribution.}
    \label{fig:SI_Plot_17}
\end{figure}

\section{Enabling Discovery of Anisotropic and Topological Materials}
The dielectric tensor has been extensively studied, with numerous models developed to investigate its anisotropic behavior in materials, as outlined in the main text. While continuous optical spectra have received comparatively less attention, GNNOpt conducted a comprehensive study in this area. However, as noted in their work~\cite{https://doi.org/10.1002/adma.202409175}, the anisotropic component of the predicted response is entirely absent. To address this limitation, we extend the concept of generalized optical weight beyond the isotropic (trace) component to include the full anisotropic tensor structure. The generalized optical weight is defined through the following expressions~\cite{ghoshProbingQuantumGeometry2024, PhysRevX.14.011052}:
\begin{equation}
\int_0^{\infty} \mathrm{d}\omega \, \frac{\operatorname{Re} \sigma^{xx}(\omega)}{\omega} = \frac{e^2}{2\hbar} K^{xx},
\end{equation}
where \( K^{xx} \) is the quantum weight, a geometric property of the insulating ground state. While the anisotropic part can be defined as 
\begin{equation}
\int_0^{\infty} \mathrm{d}\omega \, \frac{\operatorname{Im} \sigma^{xy}(\omega)}{\omega} = -\frac{e^2}{4\hbar} C^{xy},
\end{equation}
where \( C^{xy} \) is the Chern number. These quantities are directly linked to magneto-optical phenomena such as the Kerr effect (MOKE) and magnetic circular dichroism (MCD), which are widely used to probe electronic band structures, Berry curvature, and symmetry breaking in materials. MOKE is commonly employed in the study of magnetic ordering and domain imaging, while MCD provides insight into spin-resolved optical transitions and chiral electronic states.

When magnetic effects are considered, the off-diagonal components are naturally induced, breaking the usual $\varepsilon^{ij} = \varepsilon^{ji}$ symmetry. Our model can be straightforwardly adapted to lift this constraint, allowing for the prediction of the full \(3 \times 3\) tensor. The additional three antisymmetric channels then naturally capture magneto-optical contributions (e.g., $\varepsilon^{xy} \neq \varepsilon^{yx}$), without requiring further modification of the framework. In fact, these antisymmetric terms can be mapped directly onto spherical tensor components, for example:
\begin{equation}
    Y_{1,-1}=\frac{1}{\sqrt{2}}\left(\varepsilon^{z x}-\varepsilon^{x z}\right), \quad 
    Y_{1,0}=\frac{1}{\sqrt{2}}\left(\varepsilon^{x y}-\varepsilon^{y x}\right), \quad 
    Y_{1,1}=\frac{1}{\sqrt{2}}\left(\varepsilon^{y z}-\varepsilon^{z y}\right),
\end{equation}
which explicitly shows how the antisymmetric part of the tensor contributes to the magneto-optical response. In the \texttt{e3nn} notation, this corresponds to the additional channel structure $N_\omega (0e + 1o + 2e)$, giving a total of \(1 + 3 + 5 = 9\) components per frequency. Their connection to the quantum metric and Chern number makes them powerful tools for investigating the geometric and topological properties of solids, especially in low-symmetry and strongly anisotropic systems.

In our dataset, no external magnetic field is applied during the optical response calculations. As a result, off-diagonal conductivity components such as \(\sigma^{xy}\) are symmetry-forbidden in most crystal systems--particularly those with \(C_3\) rotational symmetry--unless time-reversal symmetry is explicitly broken. Consequently, only triclinic systems in our dataset exhibit non-vanishing off-diagonal components. For these cases, we are still able to compute the Chern number from \(\sigma^{xy}\), demonstrating the model's ability to capture tensorial anisotropy and its sensitivity to underlying geometric and topological features.

\begin{figure}[H]
    \centering
    \captionsetup{labelformat=empty}
    \includegraphics[width=0.8\linewidth]{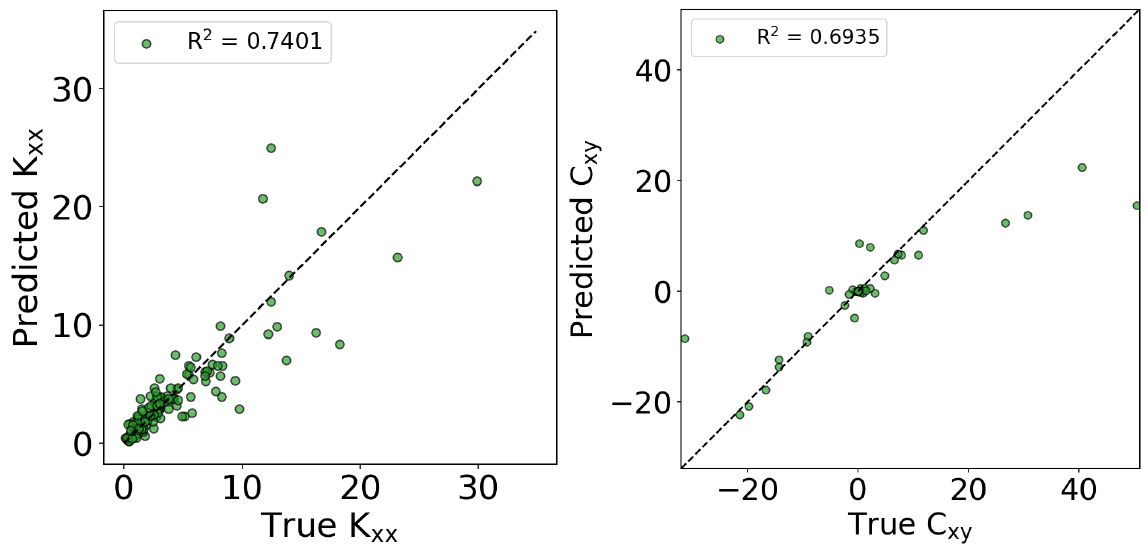}
    \caption{\textbf{Supplementary Figure 18.} Predicted versus true values for the quantum metric weight \( K^{xx} \) (left) and the Chern number \( C^{xy} \) (right) for the test set. The dashed lines indicate perfect agreement. The \( R^2 \) is reported in each panel, demonstrating the model's ability to accurately capture both the continuous optical spectra and the underlying tensorial structure.}
    \label{fig:SI_Plot_18}
\end{figure}

We emphasize the key challenges in predicting physically consistent optical spectra. As discussed in the main text, the predicted spectra must satisfy the $f$-sum rule and converge to zero in the limit (\(\omega \rightarrow \infty\)), which demands accurate modeling of the full frequency-dependent response. The Chern number \( C^{xy} \), in particular, is especially challenging to learn, as it encodes both the spectral weight distribution and the directional anisotropy arising from broken symmetries. This is a key distinction between our work and GNNOpt: our model is able to jointly and efficiently capture both the anisotropic tensorial structure and the continuous spectral behavior.

In summary, despite the absence of magnetic fields in our dataset, our model demonstrates strong capability in learning anisotropic optical responses and extracting quantum geometric quantities directly from spectral data. Looking forward, we anticipate that this framework could be extended to predict measurable MOKE and MCD responses, provided that future datasets incorporate external magnetic fields. Such an extension would significantly broaden the utility of our approach for high-throughput discovery of materials with magneto-optical functionalities.

\section{Towards Prediction of Higher-Order Tensors}
The ablation analysis discussed above (SI Section \ref{ablation_study}) shows clearly that our model can effectively capture not only the continuous spectra, but it also preserves symmetry-allowed tensor components. An interesting extension would be to consider spectra involving beyond rank-2 tensor responses such as the rank-3 shift-current conductivity $\sigma^{i j k}$, which is directly related to the bulk photovoltaic effect in materials. Shift current is a second order optical response of a material defined as: $j_i(\omega)=\sigma^{i j k}(\omega) E_j(\omega) E_k(-\omega)$, where $E_j$ and $E_k$ are electric-field components, so that $\sigma^{i j k}=\sigma^{i k j}$, which reduces the total number the independent components of $\sigma^{i j k}$ from 27 to 18. By decomposing these components into irreducible representations, one obtains contributions from three distinct angular momentum channels: $2 \times 1 o+1 \times 2 o+1 \times 3 o$. Since the shift current is odd under spatial inversion, parity must be incorporated into all irreducible representations in the network to ensure that tensor products are handled correctly.

Our model is designed to be $\mathrm{O}(3)$-equivariant, and it naturally incorporates both rotational symmetry and parity. Therefore, to predict the full shift current tensor while preserving its symmetry properties, we can simply set the output irreps to $2 N_\omega \times 1 o+N_\omega \times 2 o+N_\omega \times 3 o$, where $N_\omega$ denotes the number of grid points along the photon energy axis and controls the resolution of the continuous spectrum. We anticipate that this approach will enable accurate and comprehensive symmetry-consistent predictions for rapid and reliable identification of materials with strong shift-current responses and accelerate the discovery of promising candidates for next generation photovoltaic and optoelectronic applications.
